\documentclass[manuscript]{aastex}

\shorttitle{Abundance Distributions in Galaxies}
\shortauthors{Pilyugin et al.}

\begin{document}


\title{The Abundance Properties of Nearby Late-Type Galaxies. \\ 
II. The Relation between Abundance Distributions and Surface 
Brightness Profiles}


\author{L. S. Pilyugin\altaffilmark{1} and E. K. Grebel and I. A. Zinchenko\altaffilmark{1}}
\affil{Astronomisches Rechen-Institut, Zentrum f\"{u}r Astronomie 
           der Universit\"{a}t Heidelberg, 
           M\"{o}nchhofstr.\ 12--14, 69120 Heidelberg, Germany}
\email{pilyugin@mao.kiev.ua, grebel@ari.uni-heidelberg.de, zinchenko@mao.kiev.ua}

\author{A. Y. Kniazev\altaffilmark{2,3}}
\affil{South African Astronomical Observatory, PO Box 9, 7935 Observatory, Cape Town, South Africa}
\email{akniazev@saao.ac.za}




\altaffiltext{1}{Visiting Astronomer, Main Astronomical Observatory
            of National Academy of Sciences of Ukraine,
            27 Zabolotnogo str., 03680 Kiev, Ukraine.} 
\altaffiltext{2}{Southern African Large Telescope Foundation, PO Box 9, 7935 Observatory, Cape Town, South Africa.}
\altaffiltext{3}{Sternberg Astronomical Institute, Lomonosov Moscow State University, Moscow 119992, Russia}

\begin{abstract}
The relations between oxygen abundance and disk surface brightness (OH--$SB$ relation) 
in the infrared $W1$ band are examined for a nearby late-type galaxies. The oxygen 
abundances were presented in Paper I.  The photometric characteristics of the 
disks are inferred here using photometric maps from the literature through 
bulge-disk decomposition.  We find evidence that the OH -- $SB$ relation 
is not unique but depends on the galactocentric distance $r$ (taken as a fraction 
of the optical radius $R_{25}$) and on the properties of a galaxy:  the disk scale 
length $h$ and the morphological $T$-type. We suggest a general, four-dimensional 
OH -- $SB$ relation with the values $r$, $h$,  and $T$ as parameters. The parametric 
OH -- $SB$ relation  reproduces the observed data better than a simple, one-parameter relation; 
the deviations resulting when using our parametric relation are smaller by a 
factor of $\sim$1.4 than that the simple relation. 
The influence of the parameters on the OH -- $SB$ relation  varies with galactocentric 
distance. The influence of the $T$-type on the OH -- $SB$ relation 
is negligible at the centers of galaxies and increases with galactocentric 
distance.  In contrast, the influence of the disk scale length on the 
OH -- $SB$ relation is maximum  at the centers of galaxies and decreases 
with galactocentric distance, disappearing at the optical edges of galaxies. 
Two-dimensional relations can be used to reproduce the observed 
data at the optical edges of the disks and at the centers of the disks. 
The disk scale length should be used as a second parameter in the OH -- $SB$ 
relation at the center of the disk while the morphological $T$-type should 
be used as a second parameter in the relation at optical edge of the disk.  
The relations between oxygen abundance and disk surface brightness in 
the optical $B$ and infrared $K$ bands at the center of the
disk and at optical edge of the disk are also considered.  
The general properties of the abundance -- surface brightness relations are 
similar for the three considered bands $B$, $K$, and $W1$.  

\end{abstract}





\section{Introduction}

The chemical properties of late-type galaxies at the present epoch are
described by two values: the gas-phase abundance at a given
(predetermined) galactocentric distance (characteristic abundance) and
the radial abundance gradient.  The value of the oxygen abundance at
the $B$-band effective (or half-light) radius of the disk
\citep{Garnett1987ApJ317,Garnett2002ApJ581}, the value of the central
oxygen abundance extrapolated to zero radius from the radial abundance
gradient \citep{VilaCostas1992MNRAS259}, the value of the oxygen
abundance at $r=0.4R_{\rm 25}$, where $R_{\rm 25}$ is the isophotal
(or photometric) radius \citep{Zaritsky1994ApJ420}, and the value of
the oxygen abundance at one disk scale length from the nucleus
\citep{Garnett1997ApJ489} all have been used as the characteristic
oxygen abundance in a galaxy.  The slope of the abundance gradient is
usually expressed in terms of dex $R_{\rm 25}^{-1}$ or in terms of dex
kpc$^{-1}$.

The correlations between the characteristic oxygen abundance, the
radial abundance gradient, and global, macroscopic properties (such as
luminosity, stellar mass, Hubble type, rotation velocity) of spiral
and/or irregular galaxies were the subject of many investigations.
\citet{Lequeux1979AA80} were the first who revealed that the oxygen
abundance correlates with total galaxy mass for irregular galaxies, in
the sense that the higher the total mass, the higher the heavy element
content. The existence of correlations between the characteristic
oxygen abundance and the luminosity (stellar mass, Hubble type,
rotation velocity) of nearby late-type (spiral and irregular) galaxies
was found by
\citet{VilaCostas1992MNRAS259,Zaritsky1994ApJ420,Pilyugin2004AA425,
Pilyugin2007MNRAS376,Pilyugin2007ApJ669,Moustakas2010ApJS190},
and \citet{Berg2012ApJ754},
among many others.  

The amount of available spectra of emission-line galaxies has
increased significantly because of the completion of several large
spectral surveys, e.g., the Sloan Digital Sky Survey (SDSS)
\citep{York2000AJ120}.  Those measurements are used for abundance
determinations that provide the extended basis for investigations of
the mass (luminosity) -- metallicity relation. The existence of the
correlation between the characteristic oxygen abundance and the
luminosity (stellar mass) was confirmed using many thousands of SDSS
galaxies 
\citep[e.g.,][]{Kniazev2003ApJ593,Kniazev2004ApJS153,Tremonti2004ApJ613,Thuan2010ApJ712}. 

In contrast to the behavior of the characteristic abundance, the slope
of the radial abundance gradients does not significantly correlate
with the global properties of galaxies
\citep{VilaCostas1992MNRAS259,Zaritsky1994ApJ420,Sanchez2014aph,Paper1}.  According
to \citet{Zaritsky1994ApJ420}, the lack of a correlation between
gradients and global properties of late-type galaxies may suggest that
the relationship between these parameters is more complex than a
simple correlation. Indeed, \citet{VilaCostas1992MNRAS259} have
concluded that a correlation is seen for non-barred galaxies. 

The only really significant result for understanding the origin of the
radial abundance gradients in the disks of late-type galaxies is the
correlation found between the local oxygen abundance and the stellar
surface brightness or surface mass density
\citep{Webster1983MNRAS204,Edmunds1984MNRAS211,VilaCostas1992MNRAS259,
Ryder1995ApJ444,Moran2012ApJ745,RosalesOrtegaetal2012ApJ756,Sanchez2014aph}.
The maximum difference in oxygen abundances of H\,{\sc ii} regions at 
similar local stellar surface brightnesses in different galaxies can be as 
large as $\sim 0.5$~dex \citep[see, e.g., Fig.\ 9 in][as well our data 
below]{Ryder1995ApJ444}. 
The maximum difference in oxygen abundances of H\,{\sc ii} regions at similar
local surface mass densities is lower, $\sim 0.4$~dex, 
\citep[Fig.\ 1 in][]{RosalesOrtegaetal2012ApJ756} with a 1$\sigma$ scatter 
of the data about the median of $\pm$0.14 dex.  
We consider this
as a hint that the simple ordinary relationship between the abundance
and stellar surface brightness can be only a rough approximation and
the dependence can vary appreciably both with galactocentric distance
within a given galaxy as well as from galaxy to galaxy. In other
words, one can expect that a parametric relationship between abundance
and surface brightness reproduces the observed data better than the
ordinary relation.

In this study we will examine whether the dependence between the
abundance and stellar surface brightness varies with galactocentric
distance within a given galaxy and from galaxy to galaxy as well as
which parameters control those variations.  To answer these questions
we will examine the relations between the abundance and stellar
surface brightness at different galactocentric distances, in  
particular at the center of the disk and at the isophotal
$R_{25}$ radius (or the optical edge) of the disk. Both simple
(one-dimensional) relationships between the abundance and stellar
surface brightness and parametric (two- and four-dimensional) relationships with
different parameters will be considered.  Our paper is
organized in the following way.  We describe the data that were used
in Section 2. We examine the abundance --  surface brightness diagrams
for different samples of galaxies in Section 3. We discuss and
summarize our results in Section 4.

\section{Data}

\subsection{Abundances}

We investigated the oxygen and nitrogen abundances in 
the disks of 130 nearby late-type galaxies in Paper I \citep{Paper1}. 
We have collected around 3740 published spectra of H\,{\sc ii} regions 
from many studies (see the list of references for the emission line flux 
measurements in Table 3 in Paper I).  Since there are different 
methods for abundance determinations being used in different works, 
the resulting abundances from these studies are not homogeneous. 
Therefore, the oxygen and nitrogen abundances in all  H\,{\sc ii} regions 
were redetermined in a uniform way. We investigated the oxygen and 
nitrogen abundance distributions across the optical disks of those 
galaxies. In particular, we find the abundances in their centers, (O/H)$_{0}$, 
and at their isophotal $R_{25}$ disk radii, (O/H)$_{R_{25}}$. 
It should be emphasized that the (O/H)$_{0}$ and (O/H)$_{R_{25}}$ values 
are not the abundances in individual 
H\,{\sc ii} regions at the corresponding galactocentric distances (in many 
galaxies we have no measurements of H\,{\sc ii} regions at the center, 
$r$ = 0, and at the optical edge of a galaxy, $r$ = $R_{25}$), 
but are determined from the fit to the radial abundance distribution. 
These (O/H)$_{0}$ and (O/H)$_{R_{25}}$ values 
form the basis of the current study.

\subsection{Surface brightness profiles in the $W1$ band}

We constructed the radial surface brightness profiles in the infrared
$W1$ band (with isophotal wavelength 3.4 $\mu$m) using the publicly
available photometric maps obtained within the framework of the {\it
Wide-field Infrared Survey Explorer (WISE)} project
\citep{Wright2010AJ140}.  The conversion of the photometric map into
the surface brightness profile is performed in several steps: \\ 
-- Extraction of the image of a galaxy from the Image Mosaic 
Service\footnote{http://hachi.ipac.caltech.edu:8080/montage/index.html}. \\
-- Interactive sky background subtraction. \\ 
-- Interactive rejection of pixels with bright stars and background galaxies. \\
-- Fitting the surface brightness by ellipses using the task 
{\sc isophote} of the package {\sc stsdas} 
   in {\sc iraf}\footnote{{\sc iraf} is distributed by the
   National Optical Astronomical Observatories, which are operated by the
   Association of Universities for Research in Astronomy, Inc., under
   cooperative agreement with the National Science Foundation.} 
   where the center of the ellipses is fixed, while the major axis 
position angle and ellipticity are free parameters. Initial
   values of the position angle and ellipticity were taken from Paper I. \\
-- Interactive determination of the mean values of the major axis
position angle and ellipticity from data of the previous step. \\
-- Derivation of the surface brightness profile using the 
task {\sc isophote} of the package {\sc stsdas} with 
   fixed ellipse center position, position angle, and ellipticity parameters. \\
In this manner we determined the surface brightness profile, position
angle, and ellipticity in the $W1$ band for each galaxy.  It should be
noted that $WISE$ images in the $W1$ band have an angular resolution
of 6.1 arcsec \citep{Wright2010AJ140}.  Therefore, on the one hand,
very small (point-like) bulges can be missed.  On the other hand,
the bulge size can be overestimated.  The $WISE$ survey in the $W1$
band is deep enough to ensure that the surface brightness
profiles extend to the optical isophotal radii $R_{25}$ and even
beyond those for many galaxies. 

All surface brightness measurements were corrected for Galactic
foreground extinction before further analysis and interpretation.  The
measurements were corrected using the $A_V$
values from the recalibration by \citet{Schlafly2011ApJ737} of the
maps of \citet{Schlegel1998ApJ500} and the extinction curve of
\citet{Cardelli1989ApJ345}, assuming a ratio of total to selective
extinction of $R_{V}$ = $A_{V}$/$E_{B-V}$ = 3.1. The $A_V$ values given
in the NASA Extragalactic Database {\sc ned} were used. We did not
attempt to correct for the intrinsic extinction of the target
galaxies. 

Surface brightness measurements in solar units were used for the analysis.
The magnitude of the Sun in the $W1$ band is obtained from its
magnitude in the $V$ band and the color of the Sun $(V-W1)_{\sun}$ =
1.608 taken from \citet{Casagrande2012ApJ761}.  The distances to the
galaxies were taken from Paper I.

\subsection{Bulge-disk decomposition}

Exponential profiles of the form 
\begin{equation}
\Sigma_{L_{d}}(r)  = (\Sigma_{L})_{0}\exp(-r/h) , 
\label{equation:disk}
\end{equation}
were used to fit the observed disk surface brightness profiles in the
$W1$ band where $(\Sigma_{L})_{0}$ is the central disk surface
brightness and $h$ the radial scale length. The bulge profiles were
fitted with a general S\'{e}rsic profile, 
\begin{equation}
\Sigma_{L_{b}}(r)  = (\Sigma_L)_{e}\exp \{-b_{n}[(r/r_{e})^{1/n} - 1]\} , 
\label{equation:bulge}
\end{equation}
where $(\Sigma_L)_{e}$ is the surface brightness at the effective
radius $r_e$, i.e., the radius that encloses 50\% of the bulge light. The
factor $b_n$ is a function of the shape parameter $n$. It can be
approximated by $b_{n}  \approx 1.9992n  - 0.3271$ for $1 < n
< 10$ \citep{Graham2001AJ121}.  Thus, the $W1$ stellar surface
brightness distribution within a galaxy was fitted with the
expression 
\begin{eqnarray}
       \begin{array}{lll}
\Sigma_L(r) & = & (\Sigma_L)_{e}\exp \{-b_{n}[(r/r_{e})^{1/n} - 1]\} \\
            & + &  (\Sigma_{L})_{0}\exp(-r/h) .                        \\
     \end{array}
\label{equation:decomp}
\end{eqnarray}
The fit via  Eq.~(\ref{equation:decomp}) will be referred to below as the pure exponential 
disk (PED) approximation.  

The parameters $(\Sigma_L)_{e}$,  $r_e$,  $n$, $(\Sigma_L)_{0}$, and  $h$ 
were determined by looking for the best fit to the observed radial surface 
brighness profile. We wish to derive a set of parameters in Eq.~(\ref{equation:decomp}) 
which minimizes the deviation $\sigma_{PED}$ of 
\begin{equation}
\sigma = \sqrt{ [\sum\limits_{j=1}^n (L(r_{j})^{cal}/L(r_{j})^{obs} - 1)^2]/n}  .
\label{equation:sigma}
\end{equation}
Here $L(r_{j})^{cal}$ is the surface brightness at the galactocentric
distance $r_{j}$ computed through the  Eq.~(\ref{equation:decomp}) and
$L(r_{j})^{obs}$ is the measured surface brightness at that
galactocentric distance. 

The obtained radial profiles of the disk components were reduced to a
face-on galaxy orientation and the bulge components were assumed to be
spherical. Note that the inclination correction is purely geometrical,
and it does not include any correction for inclination-dependent
internal obscuration.

\citet{PohlenTrujillo2006AA454} found that only around 10 -- 15\% of all
spiral galaxies have a normal/standard purely exponential disk while
the surface brightness distribution of the rest of the galaxies is
better described as a broken exponential.  Therefore, the $W1$ stellar
surface brightness distribution within a galaxy was also fitted with
the broken exponential  
\begin{eqnarray}
       \begin{array}{lll}
\Sigma_L(r) & = & (\Sigma_L)_{e}\exp \{-b_{n}[(r/r_{e})^{1/n} - 1]\} \\
            & + & (\Sigma_{L})_{0,inner}\exp(-r/h_{inner}) \;\;\;\; if \;\;\; r < R^{*} ,                        \\
            & = & (\Sigma_L)_{e}\exp \{-b_{n}[(r/r_{e})^{1/n} - 1]\} \\
            & + & (\Sigma_{L})_{0,outer}\exp(-r/h_{outer}) \;\;\;\; if \;\;\; r > R^{*}  .                        \\
     \end{array}
\label{equation:decomp2}
\end{eqnarray}
Here $R^{*}$ is the break radius, i.e., the radius at which the
exponent is changed.  The fit via  Eq.~(\ref{equation:decomp2}) will
be referred to below as the broken exponential disk (BED)
approximation.  In this case, eight parameters $(\Sigma_{L})_{e}$,
$r_{e}$,  $n$, $(\Sigma_{L})_{0,inner}$, $h_{inner}$,
$(\Sigma_{L})_{0,outer}$, $h_{outer}$, and $R^{*}$  were determined by
looking for the best fit to the observed radial surface brighness profile,
i.e., we again require that  the deviation $\sigma_{BED}$ given by
Eq.~(\ref{equation:sigma}) is minimized.

\begin{figure*}
\epsscale{0.800}
\plotone{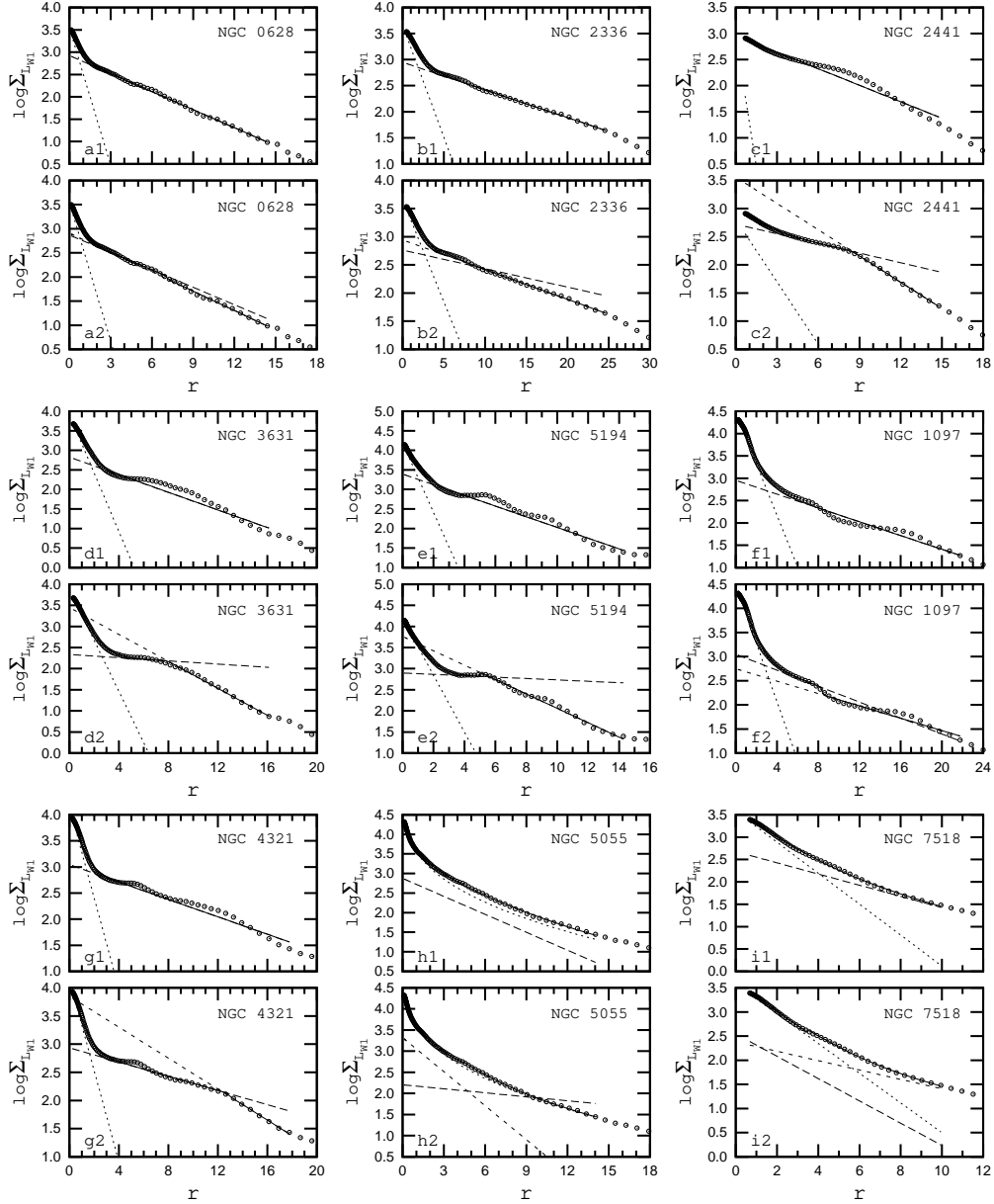}
\caption{
The resulting patterns of the bulge-disk decomposition of some of our
target galaxies. Each galaxy is presented in two panels. Each upper
panel $x1$ shows the decomposition assuming a pure exponential for the
disk (with ``$x$'' running from the letters $a$ to $i$).  The measured 
surface profile is plotted using circles. The bulge
contribution   is shown by a dotted line, the disk contribution by a
dashed line, and the total (bulge + disk) fit by a solid line.  Each
lower panel $x2$ shows the decomposition assuming a broken exponential
for the disk.  The bulge contribution is marked by a dotted line, the
inner disk by a long-dashed line, the outer disk by a short-dashed
line, and the total (bulge + disk) fit by a solid line.
\label{figure:primer}
}
\end{figure*}

Figure~\ref{figure:primer} shows the results of the bulge-disk
decomposition of some of our galaxies.  Each galaxy is displayed in two
panels. Each upper panel $x1$ shows the decomposition assuming a
pure exponential for the disk. ``$x$'' stands here for the letters $a$ 
to $i$, which are used to name the different panels in this figure.
The measured surface profile is marked
by circles.  The fit to the bulge contribution is shown by a dotted
line, the fit to the disk  by a dashed line, and the total (bulge +
disk) fitting by a solid line.  Each lower panel $x2$ shows the
decomposition assuming a broken exponential for the disk.  The bulge
contribution is shown by a dotted line, the inner disk by a
long-dashed line, the outer disk by a short-dashed line, and the total
(bulge + disk) fitting by a solid line.

Table \ref{table:samplew} lists the parameters of the surface
brightness profiles of our galaxies in the $W1$ band obtained through
the bulge-disk decomposition with the PED approximation.  The first
column contains the galaxy's name, i.e., its number in the New General
Catalogue (NGC), Index Catalogue (IC), Uppsala General Catalog of
Galaxies (UGC), or Catalogue of Principal Galaxies (PGC).  The  galaxy
inclination and the position angle of the major axis in the $W1$ band
obtained here are given in columns 2 and 3, respectively.  The
parameters of the general S\'{e}rsic profile for the bulge are listed
in columns 4 -- 6: the  logarithm of the bulge surface brightness at
the effective radius $r_{e}$  in the $W1$ band in $L_{\sun}$ pc$^{-2}$
is reported in column 4, the bulge effective radius in kpc is listed
in column 5, and the shape parameter $n$ is given in column 6.  The
logarithm of central surface brightness of the disk in the $W1$ band
in terms of $L_{\sun}$ pc$^{-2}$ is listed in column 7.  The  disk
scalelength in the $W1$ band, $h_{W1}$ in kpc, is reported in column
8.  The bulge contribution to the galaxy luminosity is reported in
column 9.  The galaxy luminosity is given in column 10.  The mean
deviation in the surface brightness $\sigma_{PED}$ around the fit
through the bulge-disk decomposition assuming a pure exponential for
the disk is given in column 11.  The mean deviation $\sigma_{BED}$
assuming a broken exponential for the disk is listed in column 12. 

\begin{figure*}
\epsscale{1.00}
\plotone{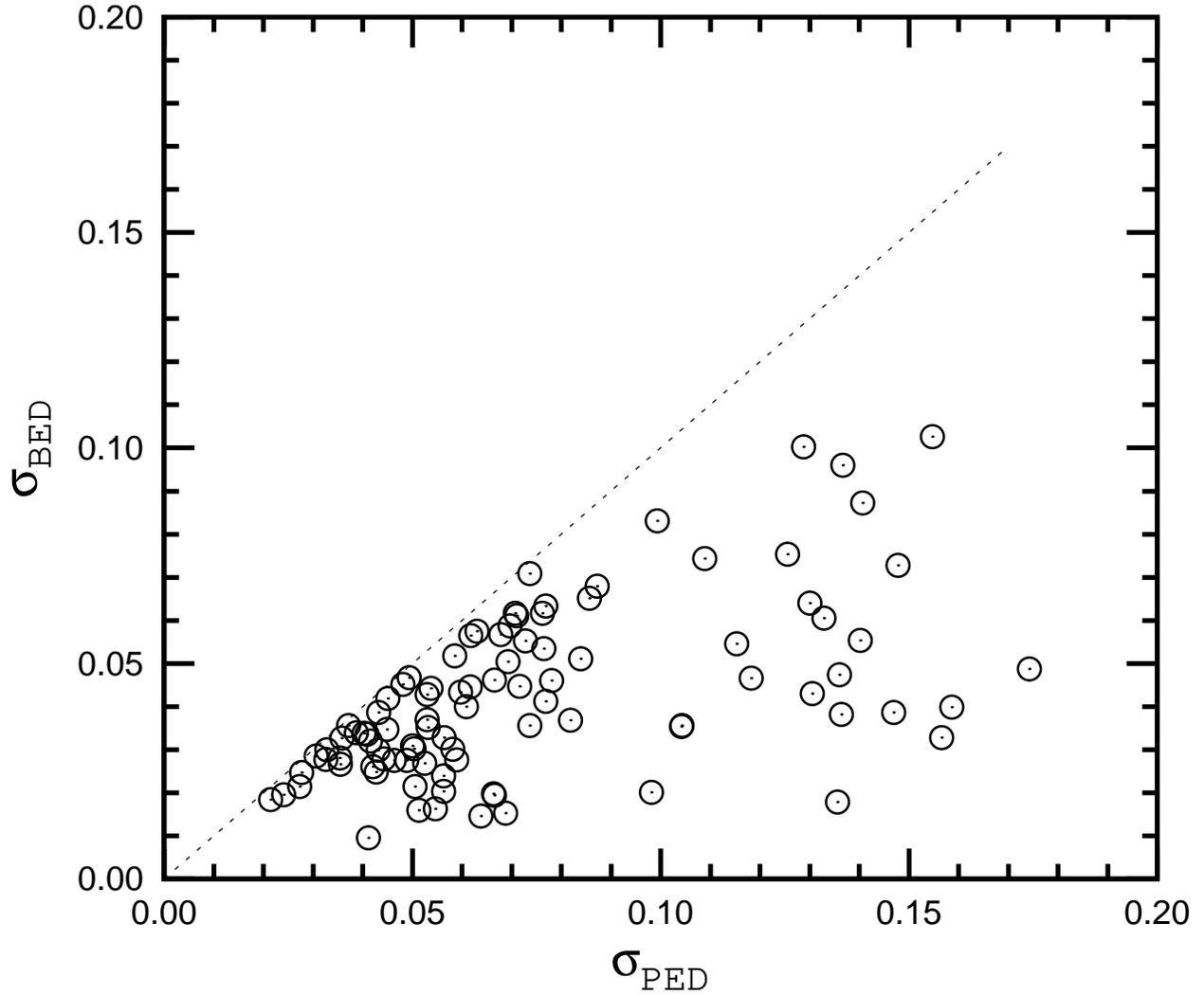}
\caption{
Quality of the surface brightness profile fit obtained via the
bulge-disk decomposition.  The mean deviation $\sigma_{BED}$ of the
fit assuming a broken exponential disk is plotted versus the mean
deviation $\sigma_{PED}$ of the fit assuming a pure exponential disk.
The dotted line shows equal values.
\label{figure:s1-s2}
}
\end{figure*}

The accuracy of the surface brightness profile fitting through the
bulge-disk decomposition assuming a pure exponential for the disk is
specified by the mean deviation $\sigma_{PED}$.  The accuracy assuming
a broken exponential for the disk is specified by the mean deviation
$\sigma_{BED}$.  Figure~\ref{figure:s1-s2} shows the comparison
between the mean deviations $\sigma_{PED}$ and  $\sigma_{BED}$.  The
surface brightness profiles of a number of galaxies are well fitted
both with the pure and with the broken exponential disks. The mean
deviation $\sigma_{PED}$ is small and close to the mean deviation
$\sigma_{BED}$.  The panels $a$ and $b$ of Figure~\ref{figure:primer}
show the surface brightness profile fitting for two of these galaxies:
NGC~628 (with $\sigma_{PED}$ = 0.041 and $\sigma_{BED}$ = 0.034) and
NGC~2336 (with $\sigma_{PED}$ = 0.045 and $\sigma_{BED}$ = 0.035).
The galaxy NGC~2336 has the largest disk scale length of around 8 kpc
among the galaxies of our sample (if the adopted distance to this
galaxy is correct).

The surface brightness profiles of a number of galaxies are much
better fitted with broken exponential disks than with pure
exponential disks.  The mean deviation $\sigma_{PED}$ is appreciably
larger than the mean deviation $\sigma_{BED}$.  Panels $c$, $d$, and
$e$ of Figure~\ref{figure:primer} show the surface brightness profile
fitting for three of these galaxies: 
NGC~2441  (with $\sigma_{PED}$ = 0.136 and $\sigma_{BED}$ = 0.018),  
NGC~3631  (with $\sigma_{PED}$ = 0.174 and $\sigma_{BED}$ = 0.049), and   
NGC~5194  (with $\sigma_{PED}$ = 0.129 and $\sigma_{BED}$ = 0.055).   
The galaxy NGC~3631 has the largest value of the mean deviation
$\sigma_{PED}$.  Galaxies with strongly broken exponential disks
(with a large difference between the disk scale lengths inside and
otside the break point) belong to this group of galaxies.  
Galaxies with bright spiral arms starting at the ends of the bar 
also belong to this group.

The surface brightness profiles of a number of galaxies are not well
fitted with either pure exponential disks nor with broken exponential
disks.  Their mean deviations $\sigma_{PED}$ and $\sigma_{PED}$ are
large.  The panels $f$ and $g$ of Figure~\ref{figure:primer} show the
surface brightness profile fitting results for two examples of such
galaxies: 
NGC~1097  (with $\sigma_{PED}$ = 0.129 and $\sigma_{BED}$ = 0.100) and 
NGC~4321  (with $\sigma_{PED}$ = 0.126 and $\sigma_{BED}$ = 0.075).   

In several cases, we could not determine a reliable disk scale length
$h_{W1}$ and/or central surface brightness of the disk
$(\Sigma_{L_{W1}})_{0}$.  The panels $h$ and $i$ of
Figure~\ref{figure:primer} show the surface brightness profile fitting
for two examples of such galaxies: NGC~5055  and NGC~7918.  The formal
values of the mean deviations $\sigma_{PED}$ and $\sigma_{BED}$ can be
rather small: $\sigma_{PED}$ = 0.045 and $\sigma_{BED}$ = 0.038 for
NGC~5055  and and $\sigma_{PED}$ = 0.038 and $\sigma_{BED}$ = 0.028
for NGC~7918.  However, the disk contribution to the surface
brightness is close to the observed  surface brightness profile over a
small interval of radial distances (if any) only (in fact, this is a
bulge-dominated galaxy).     Therefore, the values of the disk scale
length and/or central surface brightness of the disk are questionable.
Those galaxies were thus excluded from further consideration.  

Our final list involves 95 galaxies with 
estimates of the disk scale length $h_{W1}$ and central surface
brightness of the disk $(\Sigma_{L_{W1}})_{0}$ in the $W1$ band.

\subsection{Comparison to other studies}

\begin{figure*}
\epsscale{1.00}
\plotone{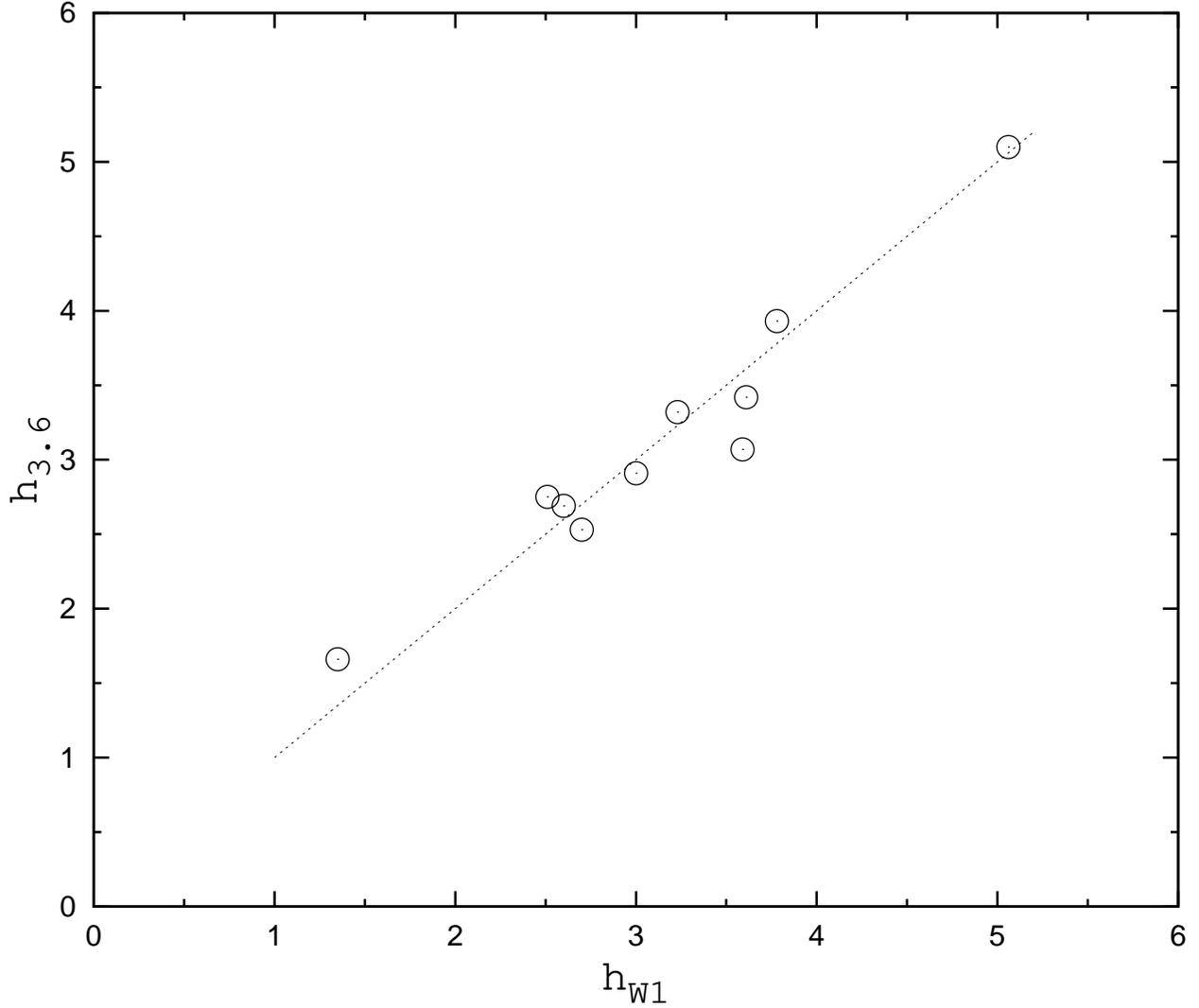}
\caption{
Comparison between disk scale lengths in the $W1$ $WISE$ band with 
the isophotal wavelength at 3.4 $\mu$m (this study) and in the 3.6 $\mu$m 
band of the {\it Spitzer Survey of Stellar Structure in Galaxies} 
\citep{Munoz2013ApJ771} (see text).  The dotted line shows equal values.
\label{figure:h-h}
}
\end{figure*}

\citet{Munoz2013ApJ771} have analyzed the surface brightness profiles
in a sample of nearby disk galaxies using deep $IRAC-1$ (3.6 $\mu$m)
images from the {\it Spitzer Survey of Stellar Structure in Galaxies
(S\,$^4$G)}.  There are a number of galaxies in common with our study.
Since the isophotal wavelength of the $W1$ $WISE$ band (3.4 $\mu$m) is
close to that of  $IRAC-1$ {\it S\,$^4$G} we can compare the resulting
disk scale lengths from our study and from \citet{Munoz2013ApJ771}.
Since we will use the disk scale length derived with the pure
exponential disk approximation in our further analysis we will focus
on these values.  \citet{Munoz2013ApJ771} have used both pure
exponential disks as well as broken exponential disks in their fits,
but they reported only their results for the case of the broken
exponential disk approximation, i.e., the values of the disk scale
length inside and outside of the break point.  

In a number of galaxies no disk break was detected.  For those
galaxies single disk scale lengths were reported.  There are four such
galaxies in common with our sample.  To enlarge the comparison sample
of galaxies, their inner disc scale length was considered as a global
disk scale length (and compared with our disk scale length obtained
with the PED approximation) if the galactocentric distance of the
break point exceeds the inner disk scale length by a factor of
$\sim$4.  

Figure~\ref{figure:h-h} shows the comparison between our disk scale
lengths and those from \citet{Munoz2013ApJ771} for ten galaxies.
Figure~\ref{figure:h-h} shows that the our disk scale lengths are in
satisfactory agreement (average difference $\sim$~9\%) with the ones 
from \citet{Munoz2013ApJ771}.

\subsection{Data in the $B$ and $K$ bands}

We also compile the radial surface brightness profiles in the
photometric $B$ and $K$ bands for galaxies of our sample. In some
cases we have used published surface brightness profiles (e.g., data
from
\citet{deJong1994AAS106,Jarrett2003AJ125,Munoz2009ApJ703,Li2011ApJS197}).
When only photometric maps of galaxies were available (e.g., $B$ and
$K$ photometric maps of \citet{Knapen2004AA426}, or $g$ and $r$
photometric maps of SDSS DR9 \citep{Ahn2012ApJS203}) then the radial
surface brightness profiles were determined in the way described
above.  Position angle and inclination angle of a given galaxy were
taken from Paper I and were kept fixed for isophotes at all
galactocentric distances.   The measurements in the SDSS filters $g$
and $r$ were converted  to $B$-band magnitudes, and the $AB$
magnitudes were reduced to the Vega photometric system using the
conversion relations and solar magnitudes of \citet{Blanton2007AJ133}.
Radial surface brightness profiles in the photometric $B$ and $K$
bands were found for 32 galaxies listed in Table \ref{table:samplew}.

Table \ref{table:samplebk} lists the characteristics of the disk in
the $B$ and $K$ bands for each galaxy obtained through the bulge-disk
decomposition with the PED approximation.  The first column contains
the galaxy name, i.e., its number in the New General Catalogue.  The
logarithm of the central surface brightness of the disk in the $B$
band, ($\Sigma_{L_{B}}$)$_0$ in units of $L_{\sun}$ pc$^{-2}$, is
given in column 2.  The  disk scalelength in the $B$ band, $h_B$ in
kpc, is reported in column 3.  The reference to the source for the $B$
band measurements (observed surface brightness profiles or photometric
maps of galaxies) used here for the bulge-disk decomposition, i.e.,
for the determination of the ($\Sigma_{L_{B}}$)$_0$ and $h_B$ values
(columns 2 and 3) is given in column 4.  The logarithm of the central
surface brightness of the disk in the $K$ band,
($\Sigma_{L_{K}}$)$_{0}$ in terms of $L_{\sun}$ pc$^{-2}$, is listed
in column 5.  The  disk scalelength in the $K$ band, $h_K$ in kpc, is
reported in column 6.  The reference to the source for the $K$-band
measurements used for the determination of the ($\Sigma_{L_{K}}$)$_0$
and $h_K$ (columns 5 and 6) via bulge-disk decomposition is given in
column 7.

It should be noted that in many cases reliable surface brightness
measurements in the $K$ or/and $B$ band do not extend to the isophotal
radius $R_{25}$ taken from the Third Reference Catalog ($RC3$,
\citet{RC3}). The measurements become too noisy or are missing beyond
some radius.

\section{The relation between abundance and surface 
brightness of the disk}

\subsection{Preliminary remarks}

The relation between abundance and surface brightness of the disk (or
surface mass density) was considered in a number of studies
\citep{Webster1983MNRAS204,Edmunds1984MNRAS211,VilaCostas1992MNRAS259,
Ryder1995ApJ444,Moran2012ApJ745,RosalesOrtegaetal2012ApJ756,Sanchez2014aph}.
There are two approaches to the investigation of the relation between
abundance and surface brightness in disks of galaxies (the OH -- $SB$
relation) or the relation between abundance and disk surface mass
density.  The first approach is to compare the local abundance with
the local surface brightness (local mass density).  But since the
measured surface brightness of the central part of a galaxy is the sum
of the bulge and disk brightnesses, the relation between measured
local abundance and local surface brightness cannot be interpreted as
a relation between disk parameters.  Another approach is to compare
the parameters of the radial abundance distribution with the
parameters of the surface brightness profile of the disk derived
within the framework of the adopted models. 

We follow the latter approach and adopt the simplest (single
exponential) model for the abundance and for the surface brightness
distributions across the disk.  The disk surface brightness
distribution is obtained through the bulge-disk decomposition of the
measured surface brightness of the galaxy. The advantage of this model
is that the radial distribution of metallicity and surface brightness
within the framework of the model can be specified by only two
parameters in two ways.  First, the radial distribution of the
abundance (surface brightness) can be characterized by the value of
the abundance (surface brightness) at the center of the disk and by
the radial abundance gradient (surface brightness disk scale length).
Second, the radial distribution of the abundance (surface brightness)
can be described by the values of the abundance (surface brightness)
at the center of the disk and at the optical edge of a galaxy's
$R_{25}$ isophotal radius.  The numerical values of the three
parameters (the oxygen abundance, 12+log(O/H), the surface brightness
of the disk, log$\Sigma_{L}$, and the disk scale length in kpc, $h$)
are comparable to each other within an order of magnitude.  The value
of the physical radial abundance gradient in terms of dex per kpc is
around 2 to 3 orders of magnitude smaller than the above parameters.
Therefore, it is preferable to use the second description of the
radial abundance distributions (by the values of the abundance at the
center of the disk and at the optical edge of a galaxy's $R_{25}$
isophotal radius) when investigating the relation between abundance
and surface brightness distributions.  Another strong argument in
favor of the use of the abundance at the center and at the optical
edge of the disk instead of the radial abundance gradient will be
given below. 

This approach requires that the adopted models for the abundance
distribution and for the surface brightness profile reproduce
adequately the observed distributions.  The validity of a single
exponential  distribution of the abundances  across the optical disk
(the existence of a break in the slope of abundance gradients) has
been questioned in a number of studies \citep[][among
others]{Zaritsky1992ApJL390,VilaCostas1992MNRAS259,
Scarano2013MNRAS428,Sanchez2014aph}.  On the other hand, it is well
known that the commonly used calibrations for abundance determinations
in H\,{\sc ii} regions do not work well over the whole range of
metallicities, e.g., the $R_{3}N_{2}$ calibration \citep[][and
references therein]{Marino2013AA559}.  It has been argued that the use
of such calibrations beyond the workable range of metallicities can
result in artificial bends
\citep{Pilyugin2001AA373,Pilyugin2003AA397}.  From this point of view
the existence of such bends in the slopes of radial abundance
gradients in the disks of spiral galaxies may be questioned.  

We emphasize that we only consider the abundance distribution within
the optical edge of a galaxy's $R_{25}$ isophotal radius.  Radial
abundance distributions extending beyond this isophotal radius in the
disks of some spiral galaxies have been measured recently
\citep{Bresolin2009ApJ695,Bresolinetal2012ApJ750,Goddard2011MNRAS412}
where a shallower oxygen abundance gradient in the outer disk (beyond
the isophotal radius) than in the inner disk was found. This
discontinuity in the gradient that occurs in proximity of the optical
edge of the galaxy seems to be real. 

\begin{figure*}
\epsscale{1.00}
\plotone{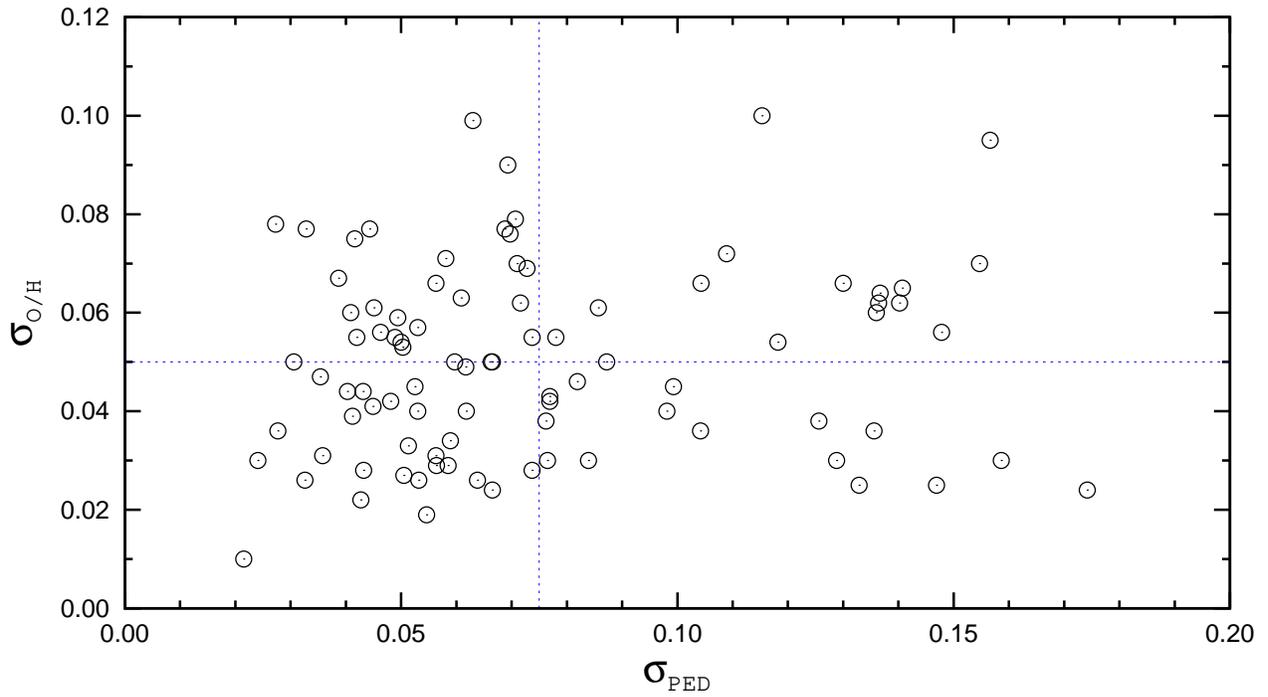}
\caption{
The deviation in oxygen abundances around the radial abundance 
gradient (in dex) versus the deviation in surface brightness 
(Eq.~(\ref{equation:sigma})) around the surface brightness profile 
fit assuming a pure exponential for the disk. Dotted lines show the average 
values of the deviations.  
\label{figure:sw-soh}
}
\end{figure*}

In Paper I the radial oxygen abundance distribution across the optical
disk in every galaxy is fitted by a single exponential relation. It
looks like a rather good approximation for the majority of the
galaxies, at least as a  first-order approximation.  Indeed, the mean
deviations in oxygen abundances around the radial abundance gradient
are usually lower than the expected precision (around 0.1 dex) of the
abundance determinations in individual H\,{\sc ii} regions (see
Figure~\ref{figure:sw-soh}).  However, a small change in slope in the
abundance distribution cannot be excluded in the disks of a number of
galaxies (e.g., in NGC~925, NGC~3184, NGC~5457).  

As mentioned above, we found that the surface brightness distributions
of disks of many galaxies can be well fitted by a pure exponential
while the surface brightness distribution of the rest of the galaxies
is better described by a broken exponential.

Figure~\ref{figure:sw-soh} shows the mean deviation in oxygen
abundances, $\sigma_{\rm O/H}$, around the radial abundance gradient
(in dex) taken from  Paper I versus the mean deviation in surface
brightness, $\sigma_{PED}$, around the surface brightness profile fit
assuming the pure exponential disk from Table \ref{table:samplew}.  If
the correlation between local values of metallicity and surface
brightness in the disk of a given galaxy is much tighter than the
correlation between the global distributions of those parameters or if
the single exponential model for the abundance and/or for the surface
brightness distributions across the disk of a given galaxy is too
uneven then one can expect that a large value of $\sigma_{\rm O/H}$
should be accompanied by a large value of $\sigma_{PED}$, i.e., there
should be correlations between the mean deviation in oxygen abundances
and the mean deviation in surface brightness.  Inspection of
Figure~\ref{figure:sw-soh} shows that there is no such correlation.
This may be considered as evidence in favor of single exponential
models for the abundance and for the surface brightness distributions
across the disks of galaxies being acceptable at least as a zero-order
approximation.  Another test of the validity of the use of the single
exponential models  for the abundance and for the surface brightness
distributions across the disks of galaxies will be presented below.

\subsection{Expected secondary parameters in the  
OH--$SB$ relationship}

We suggest  that a simple, single-parameter OH--$SB$ relationship
can be only a rough approximation and that the OH--$SB$  dependence
can vary appreciably both with galactocentric distance within a given
galaxy as well as from galaxy to galaxy. In other words, one can
expect that a parametric OH--$SB$ relationship
reproduces the observed data better than a simple linear relation.

Which parameters may be considered as possible second parameters in
the OH--$SB$ relationship?  It is believed that the gas infall rate
onto the disk decreases exponentially with time \citep[][and
references therein]{Matteucci1989MNRAS239,Pilyugin1996AA313783,
Pilyugin1996AA313792,Calura2009AA504,Pipino2013MNRAS432}.  The
timescale of gas accretion is assumed to increase with radius. This
results in an inside-out evolution of the disks of spiral galaxies
\citep{Matteucci1989MNRAS239,Munoz2011ApJ731,GonzalezDelgado2014AA562}.
In this scenario the star formation history in the disk can be
described by the expression
\begin{equation} 
SFR(t,r) \propto A \exp{-{\frac{t}{\tau(r)}}}
\label{equation:sfh} 
\end{equation} 
In this relation, $\tau$$(r)$ is the timescale of the star formation
rate at a given galactocentric distance, and $A$ is the scale factor.
The $\tau$$(r)$ value increases smoothly along the radius of the disk.
The amount of gas converted into stars at a given radius (and,
consequently, the stellar surface mass density and the astration
level) can be deduced by integrating the star formation rate over the
galaxy lifetime. In turn, the abundance is also defined by the
astration level.  Then both the radial distributions of the stellar
surface mass density and the heavy element content are governed by the
same parameters, namely $\tau$$(r)$ and $A$. This results in a
correlation between abundance and surface brightness in each
individual galaxy.  Violent events (merging, interactions) in the
evolution of a galaxy in the recent past can affect the properties of
the galaxy, e.g., strongly interacting galaxies can undergo a
flattening of their gas-phase metallicity gradient
\citep{Rupke2010ApJ723}.

Thus, the radial distributions of the stellar surface mass density and
astration level (and, consequently, abundance) are governed by the
same parameters, $\tau$$(r)$ and $A$. This results in the correlation
between those characteristics within the disk of an individual galaxy.
In particular, the local surface mass density (or surface brightness)
can serve as a surrogate indicator of the local abundance in the
individual galaxy.  Since the $\tau$$(r)$ value varies with radius one
can expect that the  OH--$SB$ dependence also changes with radius.
Since the $\tau$$(r)$ and/or $A$ values vary from one galaxy to
another this should result in differences in the OH--$SB$ dependence
for different galaxies. 

\begin{figure*}
\epsscale{0.40}
\plotone{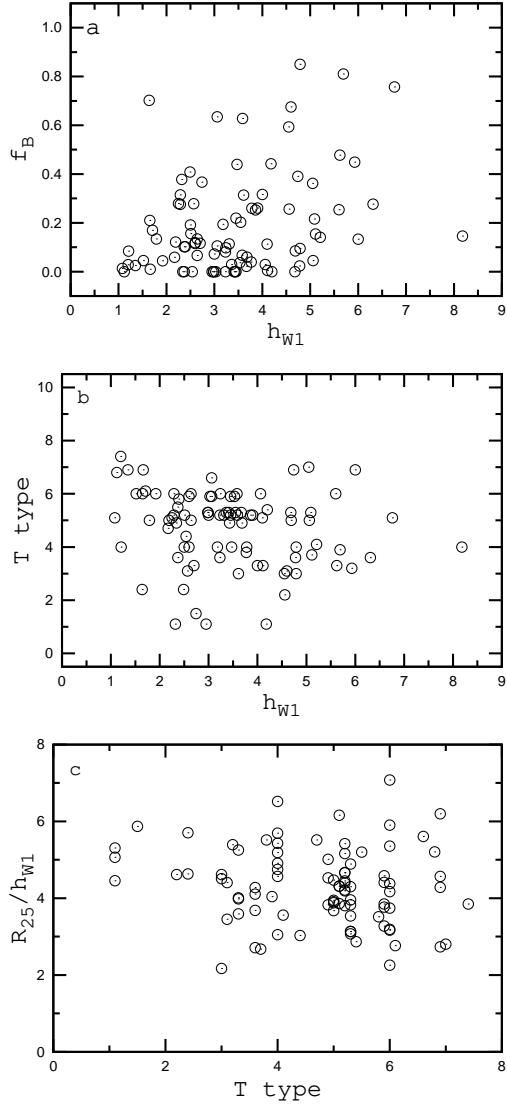}
\caption{
Bulge contribution $f_{B}$ to the galaxy luminosity in the $W1$ band  
(panel $a$) and morphological $T$ type (panel $b$) as a function of 
the disk scale length $h_{W1}$ in the $W1$ band, and ratio of disk 
radius to the disk scale length $R_{25}/h_{W1}$ as a function of morphological 
$T$ type (panel $c$).
\label{figure:h-f-t}
}
\end{figure*}

The disk scale length reflects the variation of the star formation
history with galactocentric distance. The disk scale length may then be 
considered as possible second parameter in
the  OH--$SB$ dependence. It has been known for a long
time that the morphological Hubble type of a galaxy, expressed in the
terms of $T$-type, is an indicator of the star formation history in a
galaxy \citep{Sandage1986AA161}.  Therefore, one may expect that the
morphological type is a possible second parameter in the OH--$SB$
dependence.  Furthermore, bulge stars contribute to the enrichment of
the gas in heavy elements.  Hence the bulge contribution to the galaxy
luminosity may be also considered as possible second parameter in
the  OH--$SB$ dependence, especially at the disk center.
Figure~\ref{figure:h-f-t} shows that those three parameters are
independent for our sample of galaxies.

We will examine below whether the OH--$SB$ dependence varies with
galactocentric distance within a given galaxy and from galaxy to
galaxy as well as which parameters control those variations.  To
address these questions we will examine the relations between the
abundance and stellar surface brightness at different galactocentric 
distances.  Both simple (one-dimensional) relationships between the 
abundance and stellar surface brightness and parametric (two- and 
four-dimensional) relationships with different parameters will be considered.
Firstly, the relations at the center of the disk and
at optical edge of the disk will be investigated.

\begin{figure}
\epsscale{0.30}
\plotone{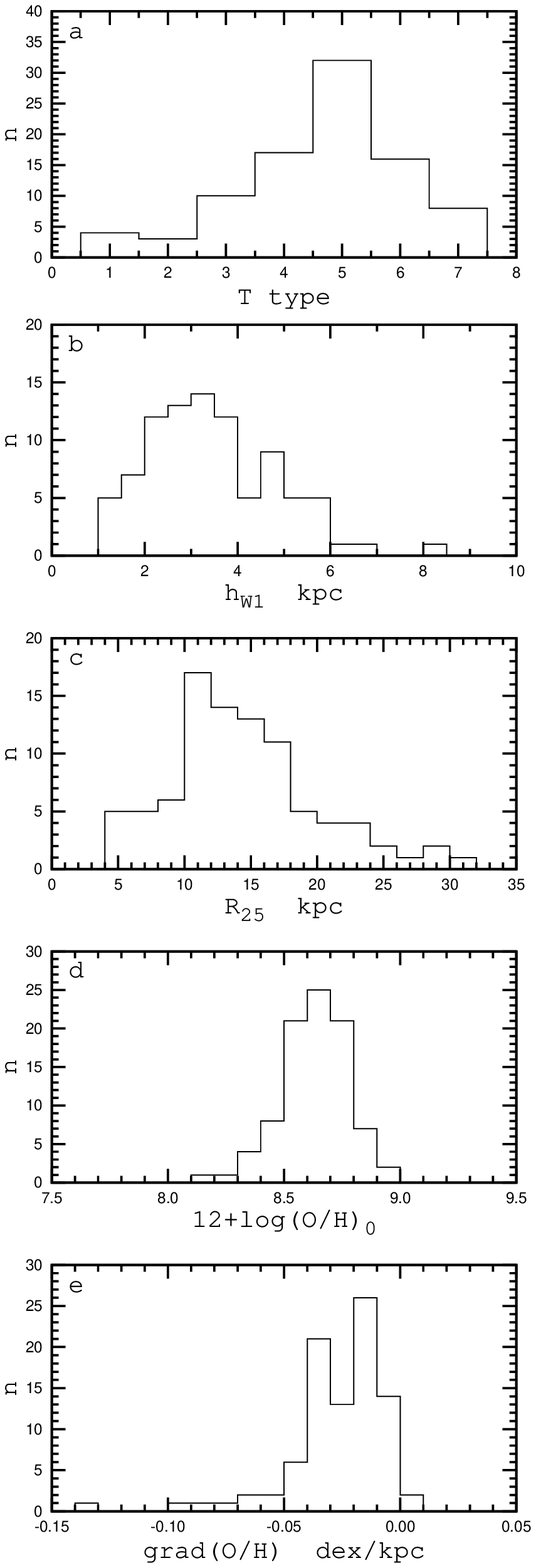}
\caption{
Histograms of morphological $T$ types (panel $a$), 
disk scale lengths in the $W1$ $WISE$ band (panel $b$),
optical radii $R_{25}$ (panel $c$), 
central oxygen abundances 12 + log(O/H)$_{0}$ (panel $d$),
and radial oxygen abundance gradients (panel $e$) for our sample of 
galaxies.
\label{figure:property}
}
\end{figure}

Here we summarize the properties of our sample of galaxies. 
Figure~\ref{figure:property} shows histograms of morphological $T$ types 
(panel $a$), disk scale lengths in the $W1$ $WISE$ band (panel $b$),
optical radii $R_{25}$ (panel $c$), central oxygen abundances 
12 + log(O/H)$_{0}$ (panel $d$), and radial oxygen abundance gradients 
(panel $e$) for our sample of galaxies. The optical radii of our galaxies 
range from $\sim$5 kpc to $\sim$30 kpc, but galaxies with radii between 
10 and 18 kpc occur most frequently. The disk scale lengths in the 
$W1$ $WISE$ band range from $\sim$~1 kpc to $\sim$~6 kpc with a few exceptions. 
The central oxygen abundances of most of the galaxies have values between   
12 + log(O/H)$_{0}$ = 8.4 and 12 + log(O/H)$_{0}$ = 8.9. The radial oxygen 
abundance gradients expressed in dex~kpc$^{-1}$ lie within the range of 
$-0.05$ to 0 although a few galaxies show steeper radial abundance gradients 
of up to $-0.1$ dex~kpc$^{-1}$ or even more. 

Panel $a$ of Figure~\ref{figure:property} shows that our sample includes 
different numbers of galaxies of different morphological types, in the sense 
that the Sc galaxies ($T$ = 5) are more numerous than the Sb ($T$ = 3) or 
 Sd ($T$ = 7) galaxies.  If the OH--$SB$ relation is dependent on the 
morphological type of a galaxy and if we will consider the single-parameter 
OH--$SB$ relation for all galaxies then the fact of unequal numbers of galaxies
of different morphological types in a sample will influence 
the result, in the sense that the derived relation will be biased towards 
the OH--$SB$ relation
for Sc galaxies. Since we will take into account the variation of the 
OH--$SB$ relation with galaxy properties (in particular, we will 
consider the parametric OH--$SB$ relation where the morphological type is a 
parameter)  then the fact of unequal numbers of galaxies of different 
morphological types in a sample will not influence the result. 
One can say that we will establish the individual relation for galaxies of 
each morphological type. Thus, it is not necessary to use any selection 
criteria in the preparation of our sample of galaxies. Therefore, we will 
consider all the galaxies with available data.
  
It can be also noted that the use of our entire sample of spiral galaxies
and of a subsample of galaxies with pure exponential disks in the
determination of 
the OH--$SB$ relation does not change the general picture (see below). It 
can be considered as indirect evidence supporting that the selection 
does not influence the result.

\subsection{The relation between central abundance and central 
surface brightness}

Our sample of the surface brightness profiles in the $W1$ band is
larger in number than that in the $B$ and $K$ bands. In addition,
these data are homogeneous both from the point of view of observations
and reduction.  Therefore we start the study of the OH--$SB$ relation
using the surface brightness in the $W1$ band. We will consider a
sample of spiral galaxies with morphological $T$-type 0.5 $\la$ $T$
$\la$ 7.5. This sample comprises 90 galaxies.

\begin{figure}
\epsscale{0.38}
\plotone{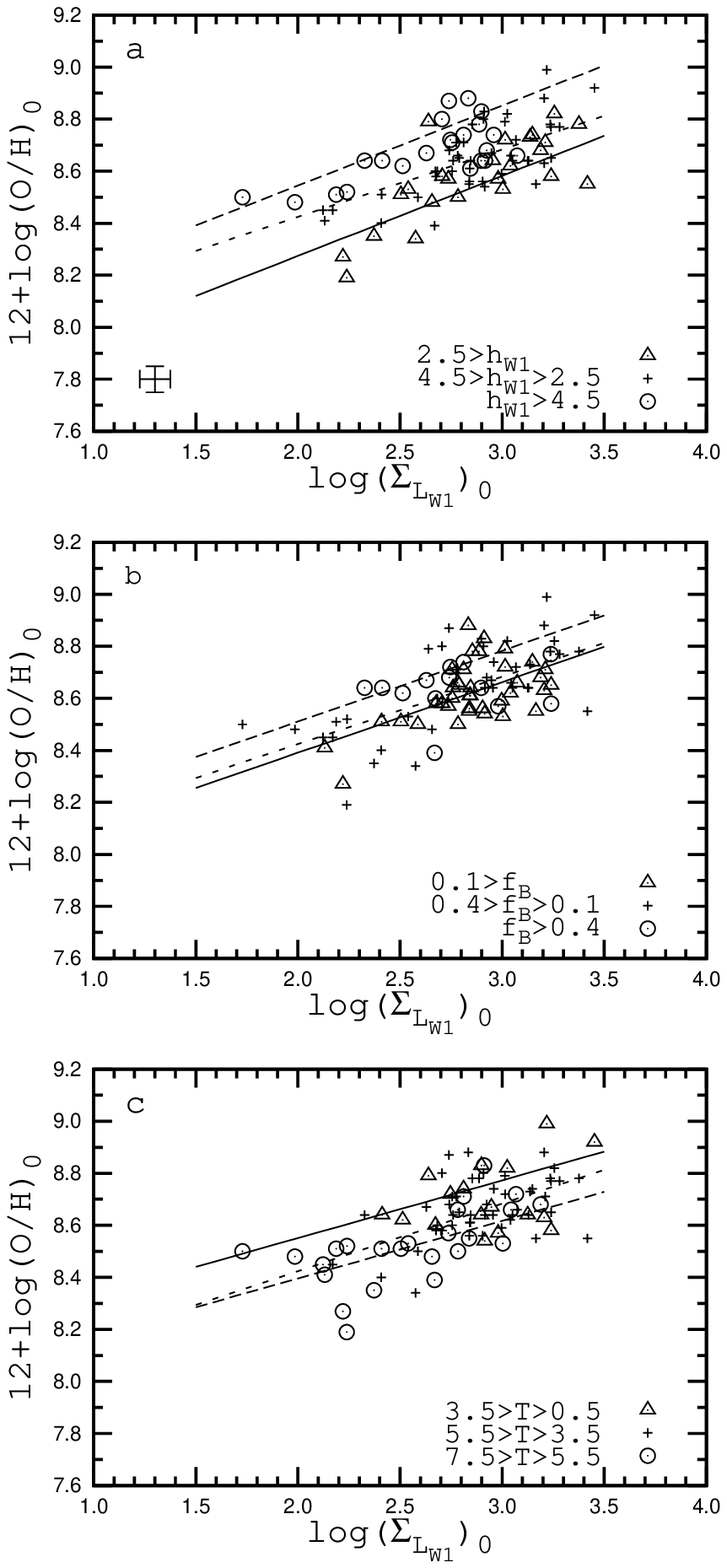}
\caption{
Central oxygen abundance as a function of central surface brightness of
the disk in the $W1$ band. The dotted line in each panel shows the
simple relation (Eq.~(\ref{equation:oho1})).  Panel $a$ shows the
subdivision of our sample of galaxies into three subsamples according
to the value of disk scale length $h_{W1}$. The solid line corresponds
to the parametric relation O/H=$f({\Sigma}_{L_{W1}},h_{W1})$
(Eq.~(\ref{equation:oho2h})) for $h_{W1}$ = 1 kpc. The dashed line is
the relation for $h_{W1}$ = 7 kpc.  
Typical errors are shown by the cross in the lower left corner.  
Panel $b$ shows the subdivision of
our sample of galaxies into three subsamples according to the value of
the bulge contribution $f_{B}$ to the total galaxy luminosity.  The solid
line corresponds the parametric relation
O/H=$f({\Sigma}_{L_{W1}},f_{B})$ (Eq.~(\ref{equation:oho2f})) for
$f_{B}$ = 0. The dashed line shows the corresponding relation  for
$f_{B}$ = 1.  Panel $c$ shows the subdivision of our sample of
galaxies into three subsamples according to the morphological $T$ type.
The solid line corresponds to the parametric relation
O/H=$f({\Sigma}_{L_{W1}},T)$  (Eq.~(\ref{equation:oho2t}))  for $T$ =
1. The dashed line represents the relation for $T$ = 7. 
\label{figure:oho}
}
\end{figure}

\begin{figure}
\epsscale{0.70}
\plotone{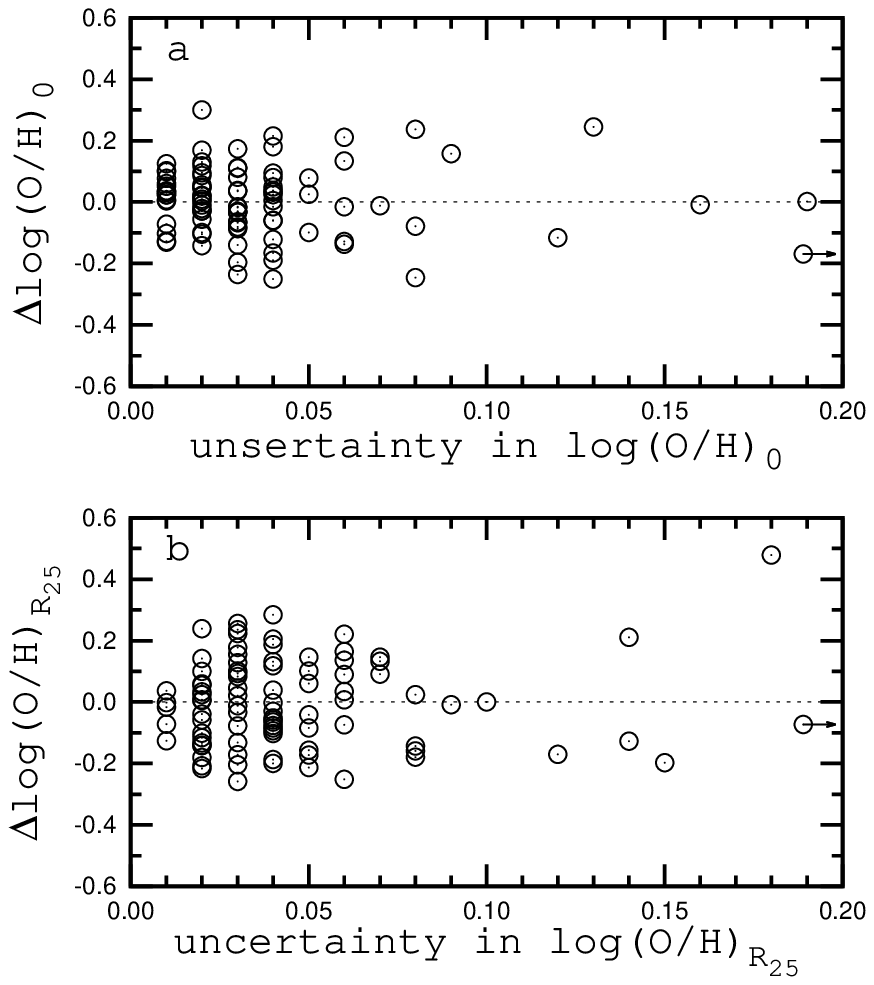}
\caption{
Panel $a$ shows deviations of the central oxygen abundances from the 
one-dimensional O/H -- $SB$ relation (the residuals of Eq.~(\ref{equation:oho1})) 
as a function of the uncertainty in central oxygen abundance. 
Panels $b$ shows deviations of the oxygen abundances at the isophotal
$R_{25}$ radius of a galaxy from the 
one-dimensional O/H -- $SB$ relation (the residuals of Eq.~(\ref{equation:ohr1})) 
as a function of the uncertainty in the oxygen abundances.  
The points show the values of the individual galaxies. 
The dotted lines indicate residuals of zero.  
}
\label{figure:err-dev}
\end{figure}

Figure~\ref{figure:oho} shows the central oxygen abundance
12+log(O/H)$_{0}$ as a function of central surface brightness of the
disk  ($\Sigma_{L_{W1}}$)$_0$ in the $W1$ band. 
The simple (one-dimensional) linear best fit relation 
\begin{equation} 12+\log
{\rm (O/H)}_{0}  = 0.259\,(\pm 0.035)\,\log (\Sigma_{L_{W1}})_{0} +
7.91 (\pm 0.10) 
\label{equation:oho1} 
\end{equation} 
is shown by the dotted line. The mean deviation around this relation 
is 0.113.  The maximum difference 
in oxygen abundances at similar local stellar surface brightnesses 
in different galaxies is as large as $\sim 0.5$~dex, similar 
to that in previous studies  \citep[e.g., Fig.~9 in][]{Ryder1995ApJ444}.
Typical errors are shown by the cross in the lower left corner. The 
typical (mean) uncertainty in the central oxygen abundances (0.04 dex) is taken from Paper I. 
The typical uncertainty in the surface brightnesses (17\% or 0.075 dex) is taken as the 
sum of the mean deviation  $\sigma_{PED}$ of the fit assuming a pure exponential disk 
and the average difference  between our disk scale
lengths and those from \citet{Munoz2013ApJ771}.
It should be noted that the error in the central oxygen abundance is more 
crucial than the error in the surface brightness. The error in the abundance 
is directly involved in the deviation of the object from the 
OH -- $SB$ relation 
while the error in the surface brightness affects the coefficient by up
to $\sim$~0.3 
(see the equations of the OH -- $SB$ relations). 

Panel $a$ in Figure~\ref{figure:err-dev} shows the residuals of
Eq.~(\ref{equation:oho1}) as a function of the uncertainty in the 
central oxygen abundances taken from Paper I.  On the one hand,
the residuals for objects with small uncertainties in the central oxygen 
abundances can be large.  On the other hand, the residuals for objects 
with large uncertainties in the central oxygen abundances can be small.  
This suggests that the deviations around the OH--$SB$ relation at the 
centers of galaxies cannot be 
attributed just to the uncertainties in the central oxygen abundances.

\begin{figure*}
\epsscale{1.00}
\plotone{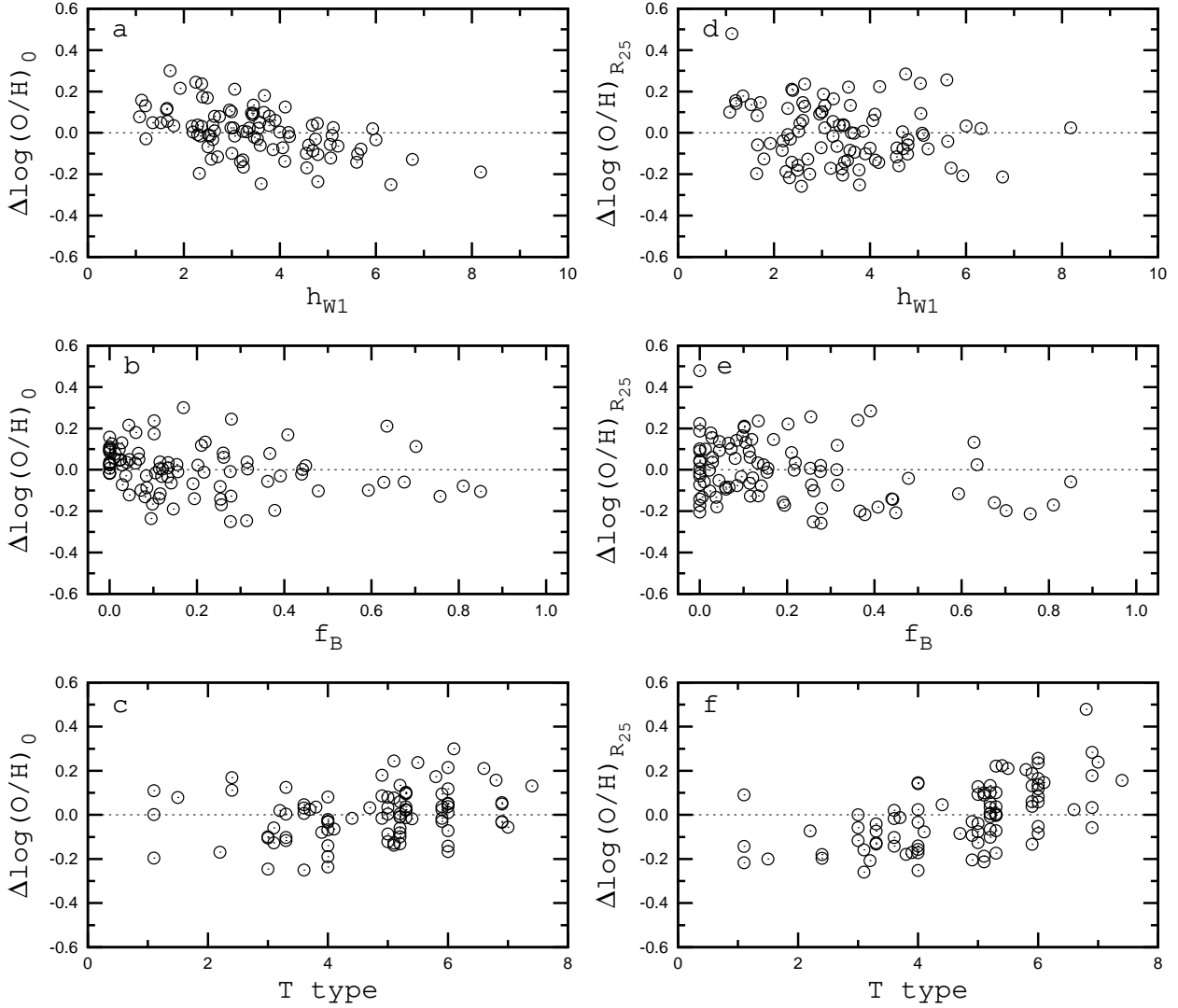}
\caption{
The residuals of Eq.~(\ref{equation:oho1}) (panels $a$, $b$, and $c$)
and Eq.~(\ref{equation:ohr1})  (panels $d$, $e$, and $f$) as a
function of the disk scale length $h_{W1}$, bulge contribution to the
total luminosity $f_{B}$, and morphological $T$ type.  The points show
the values of the individual galaxies. The dotted lines indicate 
residuals of zero. 
}
\label{figure:residual}
\end{figure*}

Figure~\ref{figure:residual} shows the residuals of
Eq.~(\ref{equation:oho1}) as a function of the disk scale length
$h_{W1}$ (panel $a$), of the bulge contribution to the total
luminosity $f_{B}$ (panel $b$), and of the  morphological $T$ type
(panel $c$).  The residuals are in excess of the typical error in the 
abundance determinations.
Figure~\ref{figure:residual} suggests that the residuals
correlate rather tightly with the disk scale length. 

Panel $a$ of Figure~\ref{figure:oho} shows the subdivision of our
sample of galaxies into three subsamples according to the value of the
disk scale length $h_{W1}$.  The parametric (two-dimensional) best-fit
relation is 
\begin{equation}
12+\log {\rm (O/H)}_{0}  = 0.308\,(\pm 0.031)\,\log (\Sigma_{L_{W1}})_{0} + 
 0.0451\,(\pm 0.0077) h + 7.61 (\pm 0.10) 
\label{equation:oho2h}
\end{equation}
The mean deviation around this relation is 0.095, i.e., it is lower than
in the case of the linear OH--$SB$ relation. The parametric OH--$SB$
relation is shown in panel $a$ of Figure~\ref{figure:oho} by the solid
line for $h_{W1}$ = 1 kpc and by the dashed line for $h_{W1}$ = 7 kpc.

Panel $b$ of Figure~\ref{figure:oho} shows the subdivision of our sample
of galaxies into three subsamples according to the value of the bulge
contribution $f_B$ to the galaxy luminosity.  The parametric best fit
relation is 
\begin{equation}
12+\log {\rm (O/H)}_{0}  = 0.272\,(\pm 0.035)\,\log (\Sigma_{L_{W1}})_{0} + 
 0.120\,(\pm 0.058) f_{B} + 7.85 (\pm 0.10) 
\label{equation:oho2f}
\end{equation}
The mean deviation about this relation is 0.110, i.e., it is close to
that for the one-dimensional relation. This two-dimensional
relation is shown in panel $b$ of Figure~\ref{figure:oho} by the solid
line for  $f_{B}$ = 0 and by the dashed line for $f_{B}$ = 1. 

Panel $c$ of Figure~\ref{figure:oho} shows the subdivision of our
sample of galaxies into three subsamples according to the
morphological $T$-type.  The two-dimensional best-fit relation is 
\begin{equation}
12+\log {\rm (O/H)}_{0}  = 0.222\,(\pm 0.036)\,\log (\Sigma_{L_{W1}})_{0} - 
 0.0258\,(\pm 0.0088) T + 8.13 (\pm 0.12) 
\label{equation:oho2t}
\end{equation}
The mean deviation about this relation is 0.107, i.e., it is close to
that for the one-dimensional relation. This two-dimensional
relation is shown in panel $c$ of Figure~\ref{figure:oho} by the solid
line for  $T$ = 1 and by the dashed line for $T$ = 7. 

We have also found the three-dimensional best fit 
\begin{eqnarray}
       \begin{array}{lll}
12+\log {\rm (O/H)}_{0} & = & 7.75\,(\pm 0.13) + 0.285\,(\pm 0.036)\,\log (\Sigma_{L_{W1}})_{0} \\  
                       & + & 0.0412\,(\pm 0.0081)\,h_{W1} - 0.0128\,(\pm 0.0082) T \\
     \end{array}
\label{equation:oho3}
\end{eqnarray}
The mean deviation around this relation is 0.094, i.e., is close to
that of two-dimensional relation O/H=$f({\Sigma}_{L_{W1}},h_{W1})$
(Eq.~(\ref{equation:oho2h})).  The two-dimensional relations
will be considered below.

\begin{figure*}
\epsscale{0.80}
\plotone{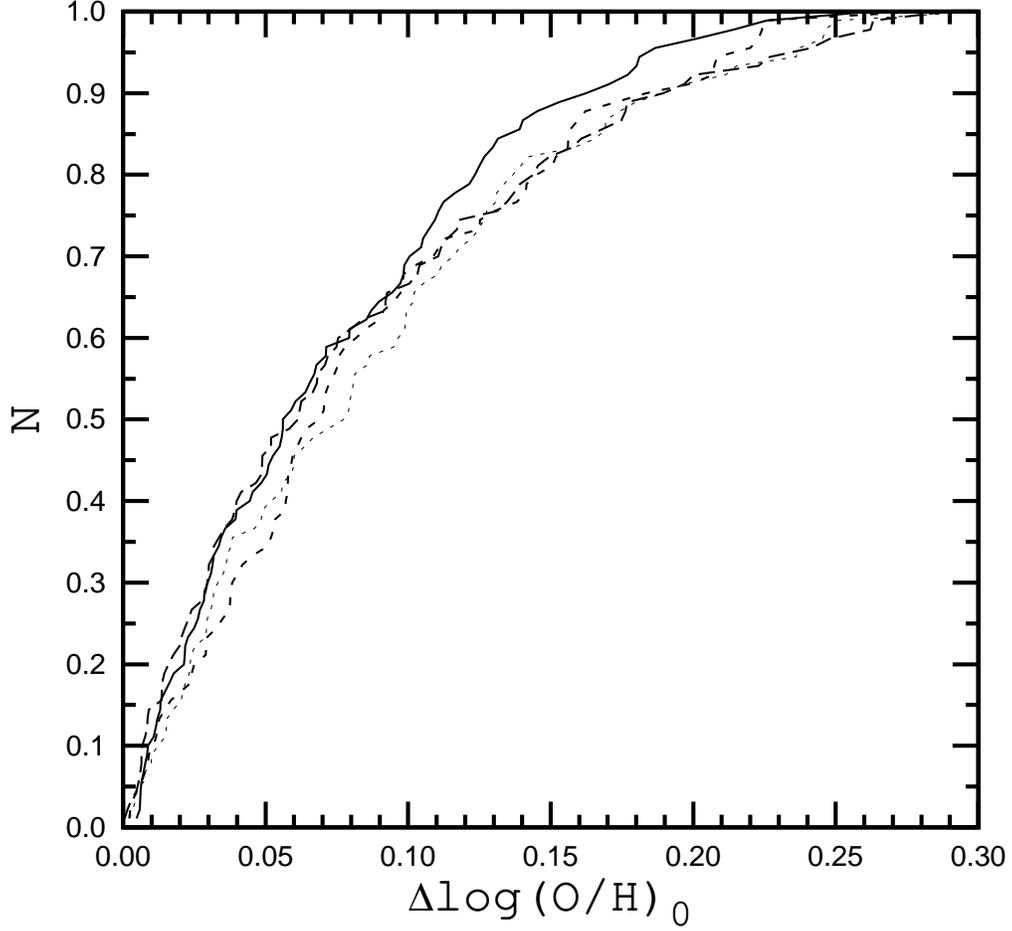}
\caption{
Cumulative number of individual galaxies with the absolute value of
the difference between observed and computed central oxygen abundance
less than a given value.  The cumulative number is normalized to the
total number of galaxies.  The computed oxygen abundances come from
the linear relation O/H=$f({\Sigma}_{L_{W1}})$
(Eq.~(\ref{equation:oho1})) (dotted line), from the parametric
relation O/H=$f({\Sigma}_{L_{W1}},h_{W1})$
(Eq.~(\ref{equation:oho2h})) (solid line), from the parametric
relation O/H=$f({\Sigma}_{L_{W1}},f_{B})$ (Eq.~(\ref{equation:oho2f}))
(long-dashed line), and from the parametric relation
O/H=$f({\Sigma}_{L_{W1}},T)$ (Eq.~(\ref{equation:oho2t}))
(short-dashed line).
\label{figure:go}
}
\end{figure*}

Thus, only the mean deviation about the parametric relation
O/H=$f({\Sigma}_{L_{W1}},h_{W1})$ is lower than that for the linear
relation.  The fact that the parametric relation
O/H=$f({\Sigma}_{L_{W1}},h_{W1})$ reproduces the observed data better
than the other relations can also be illustrated in the following way.
Figure~\ref{figure:go} shows the cumulative number of individual
galaxies with an absolute value of the difference between observed and
computed central oxygen abundance less than a given value.  The
cumulative number is normalized to the total number of galaxies.  The
computed oxygen abundances are obtained from the one-dimensional
relation O/H=$f({\Sigma}_{L_{W1}})$  (Eq.~(\ref{equation:oho1}))
(dotted line), from the two-dimensional relation
O/H=$f({\Sigma}_{L_{W1}},h_{W1})$ (Eq.~(\ref{equation:oho2h})) (solid
line), from the two-dimensional relation O/H=$f({\Sigma}_{L_{W1}},T)$
(Eq.~(\ref{equation:oho2f})) (long-dashed line), and from the
two-dimensional relation O/H=$f({\Sigma}_{L_{W1}},f_{B})$
(Eq.~(\ref{equation:oho2t})) (short-dashed line).

\begin{figure*}
\epsscale{1.00}
\plotone{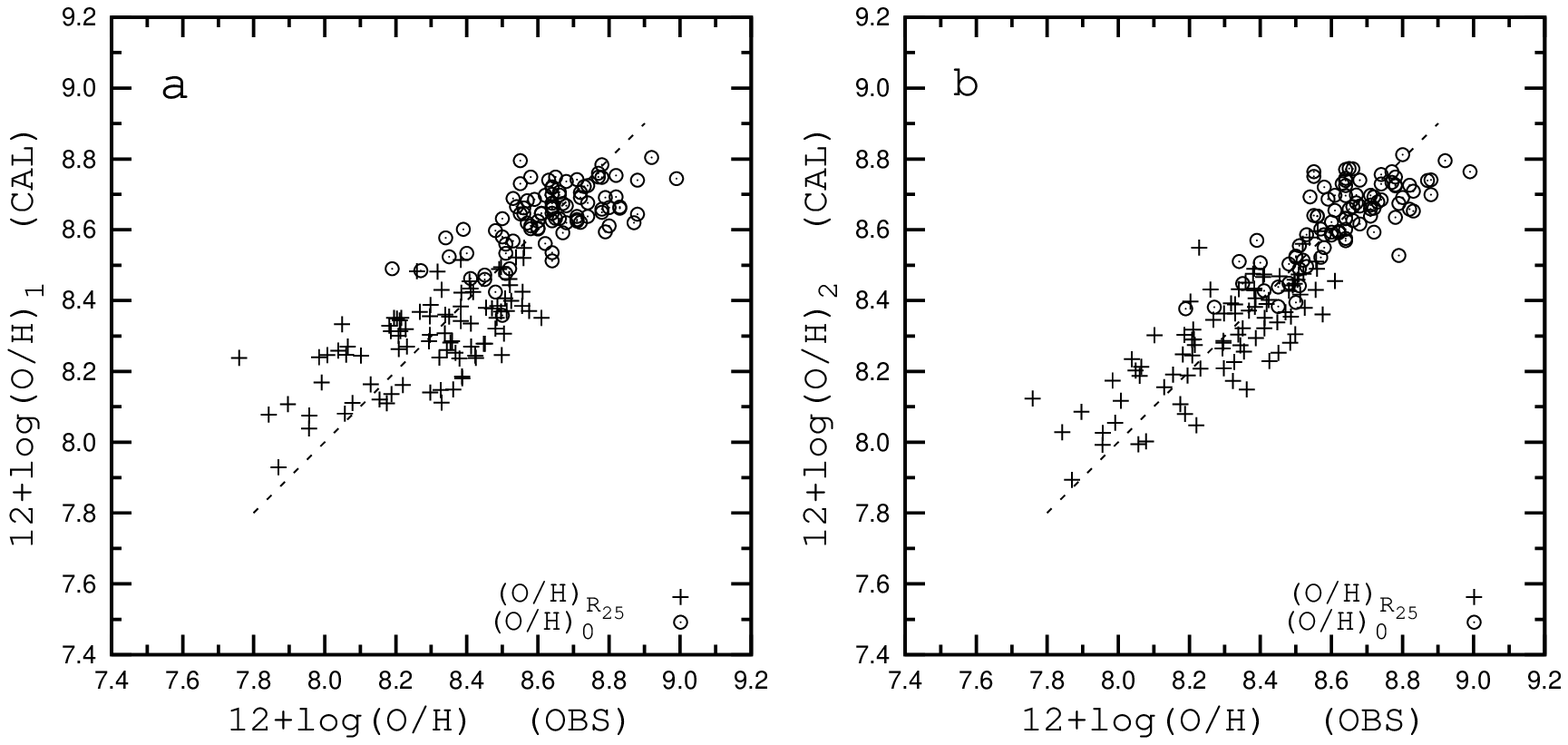}
\caption{
Comparison between computed and observed oxygen abundances at the
centers of galaxies (circles) and at the optical edge of the galaxies'
$R_{25}$ isophotal radius (plus signs). Panel $a$ shows the computed
oxygen abundances obtained from the one-dimensional relations
(Eq.~(\ref{equation:oho1}) and Eq.~(\ref{equation:ohr1})).  Panel $b$
shows the computed oxygen abundances obtained from the two-dimensional
relations  (Eq.~(\ref{equation:oho2h}) and
Eq.~(\ref{equation:ohr2t})).
\label{figure:oh-oh}
}
\end{figure*}

In Figure~\ref{figure:oh-oh} we plot the observed oxygen abundances at
the centers of galaxies (the intersect values determined from the
radial abundance distribution) versus abundances obtained from the
linear OH--$SB$ relation, Eq.~(\ref{equation:oho1}), (panel $a$) and
from the parametric relation, Eq.~(\ref{equation:oho2h}), (panel $b$).
We find that the parametric relation between central
abundance and central surface brightness of the disk with the disk
scale length as a second parameter reproduces the observed data
better than the linear OH -- $SB$ relation.

\subsection{The relation between abundance and surface 
brightness at the optical edge of a galaxy}

\begin{figure}
\epsscale{0.50}
\plotone{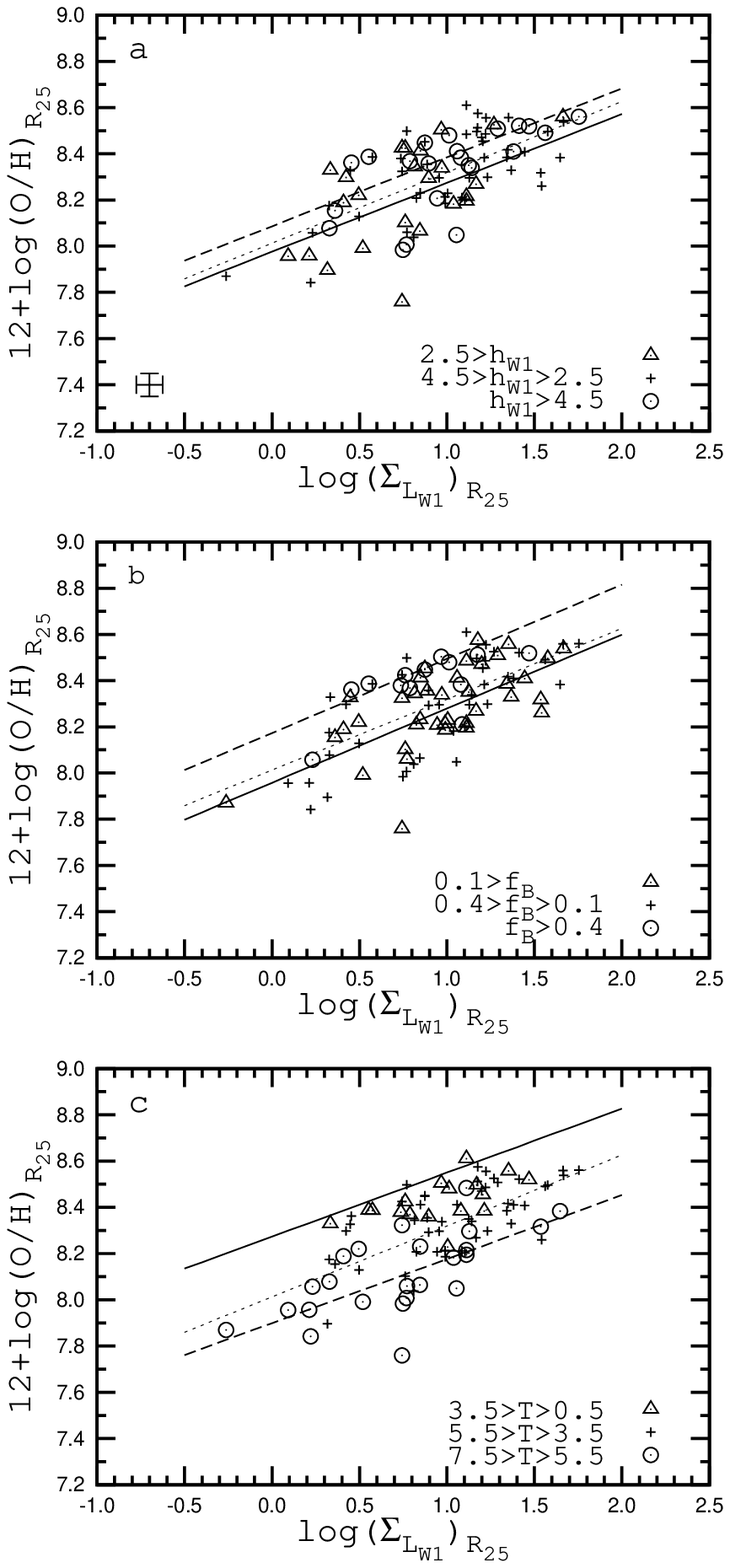}
\caption{
The same as Figure~\ref{figure:oho} but for the oxygen abundance and  
surface brightness of the disk in the $W1$ band at the optical 
edge of the galaxies' $R_{25}$ isophotal radius.
\label{figure:ohr}
}
\end{figure}

\begin{figure*}
\epsscale{0.80}
\plotone{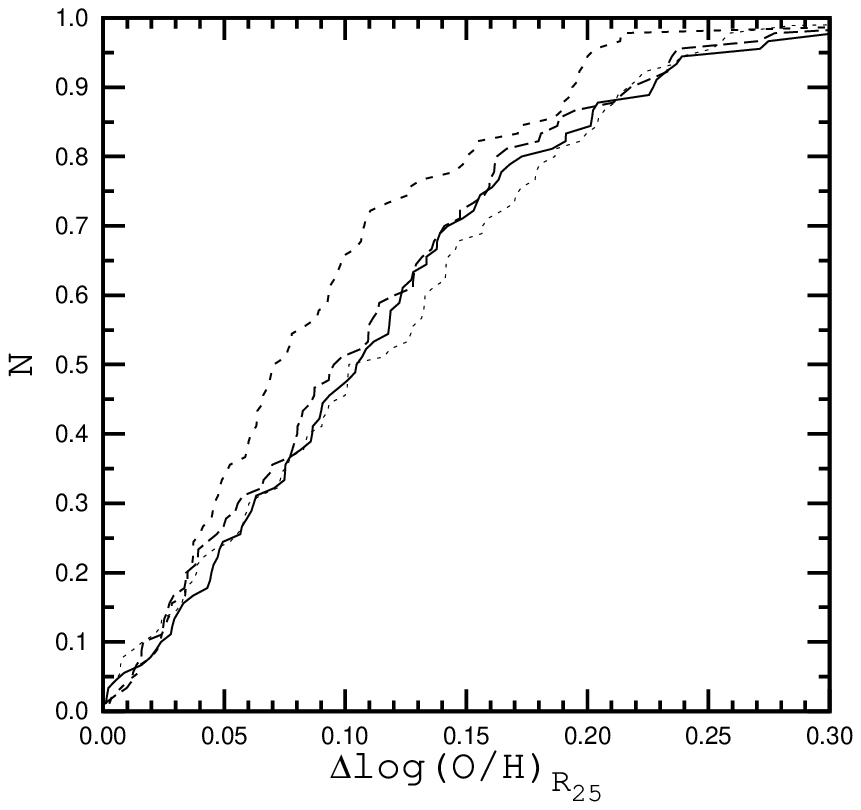}
\caption{
Cumulative number of individual galaxies for which the absolute value
of the difference between observed and computed oxygen abundance at
the optical edge of the galaxies' $R_{25}$ isophotal radius is less than
a given value.  The cumulative number is normalized to the total number of
galaxies.  The computed oxygen abundances are obtained from
the one-dimensional relation O/H=$f({\Sigma}_{L_{W1}})$
(Eq.~(\ref{equation:ohr1})) (dotted line), from the two-dimensional
relation O/H=$f({\Sigma}_{L_{W1}},h_{W1})$
(Eq.~(\ref{equation:ohr2h})) (solid line), from the two-dimensional
relation O/H=$f({\Sigma}_{L_{W1}},f_{B})$ (Eq.~(\ref{equation:ohr2f}))
(long-dashed line), and from the two-dimensional relation
O/H=$f({\Sigma}_{L_{W1}},T)$ (Eq.~(\ref{equation:ohr2t}))
(short-dashed line).
\label{figure:gr}
}
\end{figure*}

Figure~\ref{figure:ohr} shows the oxygen abundance
12+log(O/H)$_{R_{25}}$ at the optical edge of a galaxy's $R_{25}$
isophotal radius as a function of the surface brightness of the disk
($\Sigma_{L_{W1}}$)$_{R_{25}}$ in the $W1$ band. Again, the maximum difference 
in oxygen abundances at similar local stellar surface brightnesses 
in different galaxies is as large as $\sim 0.5$~dex, similar 
to that in previous studies  \citep[e.g., Fig.~9 in][]{Ryder1995ApJ444}. 
 The linear one-dimensional best fit relation 
\begin{equation}
12+\log {\rm (O/H)}_{R_{25}}  = 0.307\,(\pm 0.038)\,\log (\Sigma_{L_{W1}})_{R_{25}} + 8.01 (\pm 0.04) 
\label{equation:ohr1}
\end{equation}
is shown by the dotted line. The mean deviation around this regression
is 0.144.

Panel $b$ in Figure~\ref{figure:err-dev} shows the residuals of
Eq.~(\ref{equation:ohr1}) as a function of the uncertainty in the 
(O/H)$_{R_{25}}$ abundances. Since the deviations of objects with 
small uncertainties in the oxygen abundances can be large and, 
in contrast, the deviations of objects with large uncertainties in 
the oxygen abundances can be small  those deviations cannot be 
attributed to the uncertainties in the oxygen abundance.

Figure~\ref{figure:residual} shows the residuals of
Eq.~(\ref{equation:ohr1}) as a function of the disk scale length
$h_{W1}$ (panel $d$), of the bulge contribution to the total
luminosity $f_{B}$ (panel $e$), and of the  morphological $T$ type
(panel $f$).  Inspection of Figure~\ref{figure:residual} shows that
the residuals correlate rather tightly with the morphological $T$
type. 

Panel $a$ of Figure~\ref{figure:ohr}  shows the division of our sample of galaxies in three
subsamples according to the value of disk scale length $h_{W1}$.  The
two-dimensional best fit relation is 
\begin{equation}
12+\log {\rm (O/H)}_{R_{25}}  = 0.298\,(\pm 0.038)\,\log (\Sigma_{L_{W1}})_{R_{25}} + 
 0.0184\,(\pm 0.0112) h + 7.96 (\pm 0.05) 
\label{equation:ohr2h}
\end{equation}
The mean deviation around this relation is 0.142, i.e., it is close to
that in the case of the one-dimensional relation. The obtained
two-dimensional relation is presented in panel $a$ of
Figure~\ref{figure:ohr} by the solid line for  $h_{W1}$ = 1 kpc and by
the dashed line for  $h_{W1}$ = 7 kpc.

Panel $b$ of Figure~\ref{figure:ohr} shows the division of our sample
of galaxies in three subsamples according to the value of the bulge
contribution $f_B$ to the galaxy luminosity.  The two-dimensional best
fit relation is 
\begin{equation}
12+\log {\rm (O/H)}_{R_{25}}  = 0.321\,(\pm 0.037)\,\log (\Sigma_{L_{W1}})_{0} + 
 0.215\,(\pm 0.072) f_{B} + 7.96 (\pm 0.04) 
\label{equation:ohr2f}
\end{equation}
The mean deviation around this relation is 0.137, i.e., it is close to
the deviation of the one-dimensional relation. The obtained
two-dimensional relation is shown in panel $b$ of
Figure~\ref{figure:ohr} by the solid line for  $f_{B}$ = 0 and by the
dashed line for $f_{B}$ = 1. 

Panel $c$ of Figure~\ref{figure:ohr} shows the division of our sample
of galaxies in three subsamples according to the morphological
$T$-type.  The two-dimensional best fit relation is 
\begin{equation}
12+\log {\rm (O/H)}_{R_{25}}  = 0.277\,(\pm 0.031)\,\log (\Sigma_{L_{W1}})_{R_{25}} - 
 0.0622\,(\pm 0.0090) T + 8.34 (\pm 0.06) 
\label{equation:ohr2t}
\end{equation}
The mean deviation around this relation is 0.116, i.e., it is lower than
in the case of the one-dimensional relation. This two-dimensional
relation is shown in panel $c$ of Figure~\ref{figure:ohr} by the solid
line for  $T$ = 1 and by the dashed line for $T$ = 7. 

Figure~\ref{figure:gr} shows the cumulative number of individual
galaxies with an absolute value of the difference between observed and
computed central oxygen abundance less than a given value.  The
cumulative number is normalized to the total number of galaxies.  The
computed oxygen abundances are obtained from the one-dimensional
relation O/H=$f({\Sigma}_{L_{W1}})$  (Eq.~(\ref{equation:ohr1}))
(dotted line), from the two-dimensional relation
O/H=$f({\Sigma}_{L_{W1}},h_{W1})$ (Eq.~(\ref{equation:ohr2h})) (solid
line), from the two-dimensional relation O/H=$f({\Sigma}_{L_{W1}},T)$
(Eq.~(\ref{equation:ohr2f})) (long-dashed line), and from the
two-dimensional relation O/H=$f({\Sigma}_{L_{W1}},f_{B})$
(Eq.~(\ref{equation:ohr2t})) (short-dashed line).

In Figure~\ref{figure:oh-oh} we plot the observed oxygen abundance at
the optical edge of the disk versus the abundance obtained from the
one-dimensional relation, Eq.~(\ref{equation:ohr1}), (panel $a$) and
from the two-dimensional relations, Eq.~(\ref{equation:ohr2t}), (panel
$b$). 

Again, one of the two-dimensional relations, O/H=$f({\Sigma}_{L_{W1}},
T)$,  reproduces the observed data at the optical edge of the disk
better than the other relations.  However, the second parameter in the
relation between abundance and surface brightness at the optical edge
of the disk is not the same as the one in the relation at the center
of the disk.

\subsection{The relation between abundance and surface 
brightness as a function of galactocentric distance}

\begin{figure}
\epsscale{0.40}
\plotone{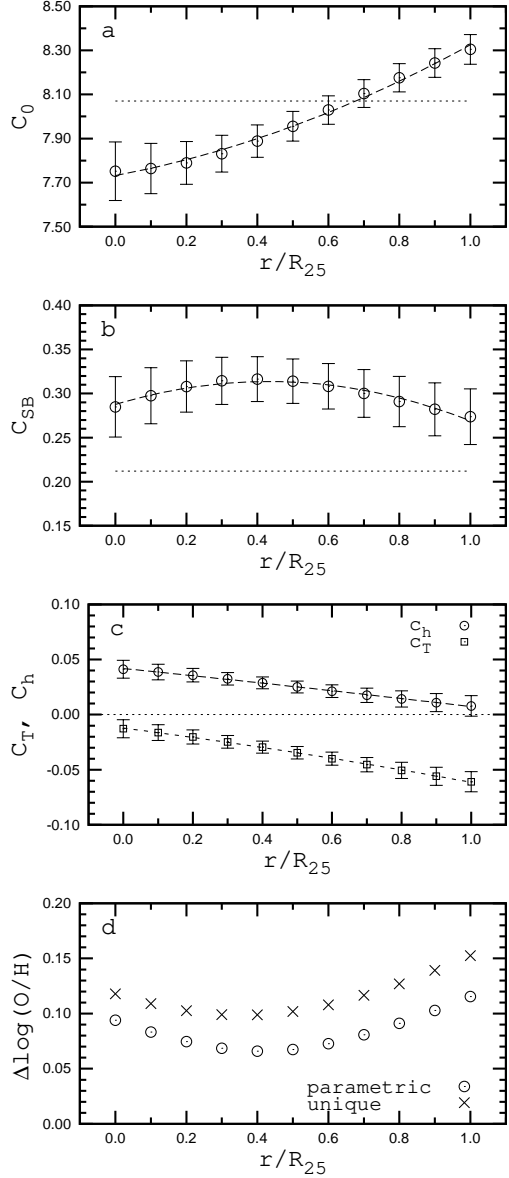}
\caption{
The coefficients $C_{0}$ (panel $a$), $C_{SB}$  (panel $b$), $C_{h}$ (panel $c$, circles) 
and $C_{T}$ (panel $c$, squares) in parametric OH -- $SB$ relation  as a function of 
galactocentric distance. The dashed lines are the best fits to the data points.
The dotted lines are the corresponding coefficients in the simple
OH -- $SB$ relation.  
Panel $d$ shows the deviations around the simple relation, residuals of Eq.~(\ref{equation:allr1}), 
(crosses) and around the parametric relation, residuals of Eq.~(\ref{equation:allr3}), (circles) 
as a function of galactocentric distance. 
\label{figure:a-r}
}
\end{figure}

\begin{figure}
\epsscale{0.500}
\plotone{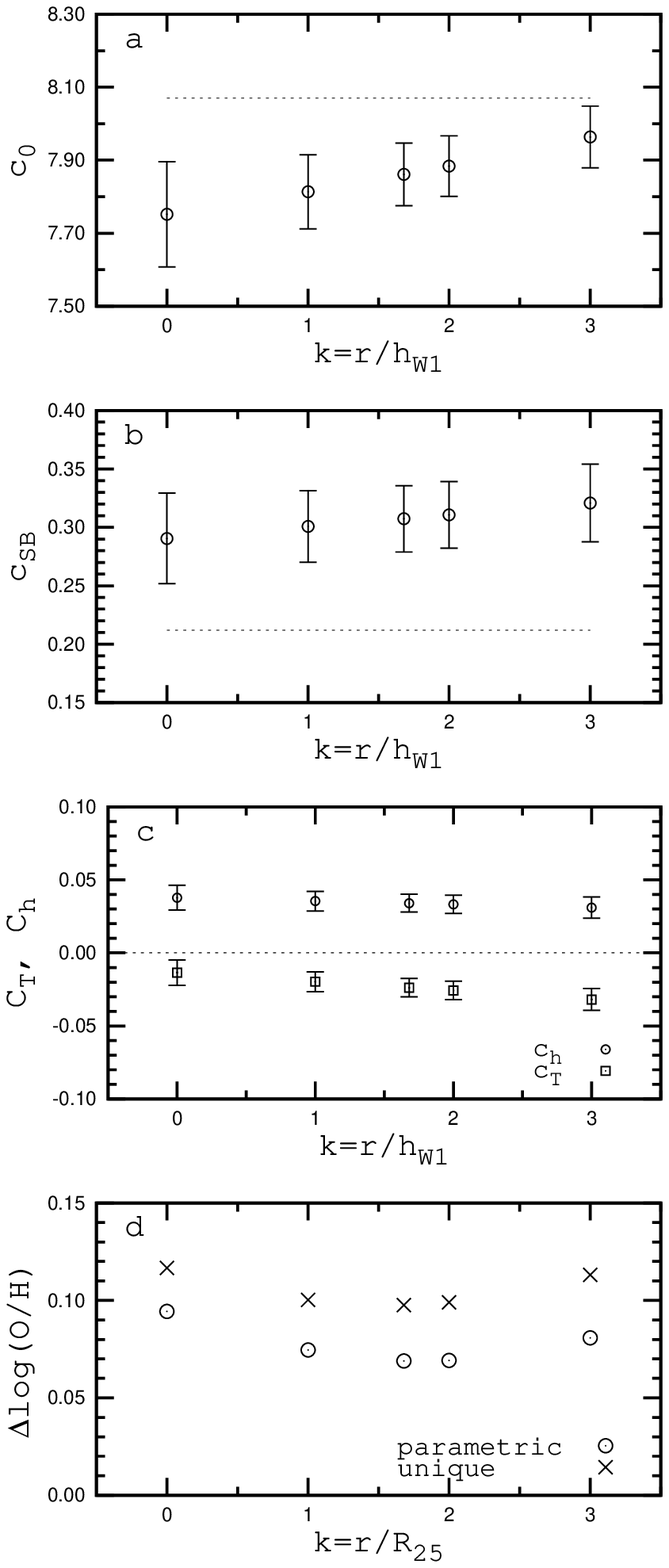}
\caption{
The same as Figure~\ref{figure:a-r} but the characteristics of each galaxy are taken at 
a galactocentric distance proportional to $k$ disk scale lengths $h_{W1}$ in 
each galaxy.
\label{figure:a-er}
}
\end{figure}

To investigate the variation of the relation between abundance and surface 
brightness across the disks of galaxies we will find the parametric O/H -- $SB$ relation at 
the different galactocentric distances expressed in terms of optical radius $R_{25}$ 
\begin{eqnarray}
       \begin{array}{lll}
12+\log {\rm (O/H)} & = & C_{0}(r)  + C_{SB}(r) \,\log (\Sigma_{L_{W1}})_{r}  \\   
                    & + & C_{h}(r)\,h_{W1} + C_{T}(r) \, T    \\
     \end{array}
\label{equation:d3}
\end{eqnarray}
with a step size of $\Delta$$r$ = 0.1$R_{25}$. 
We will also consider the simple relation 
\begin{equation}
12+\log {\rm (O/H)}  =  C_{0}  + C_{SB} \,\log (\Sigma_{L_{W1}})  . 
\label{equation:d1}
\end{equation}
The surface brightness of the disk at a given galactocentric distance is 
estimated through Eq.~(\ref{equation:disk}) using the central disk surface
brightness  $(\Sigma_{L})_{0}$  and the radial scale length $h$ from 
Table \ref{table:samplew}. 
The oxygen abundance at a given galactocentric distance is estimated from the 
central oxygen abundance and the radial abundance gradient listed in Paper I.

Figure~\ref{figure:a-r} shows the obtained  coefficients $C_{0}$ (panel $a$),  
$C_{SB}$ (panel $b$), $C_{h}$ (panel $c$, circles) and $C_{T}$ (panel $c$, squares) in the parametric 
relation as a function of galactocentric distance. 
The variation in the coefficients in  Eq.~(\ref{equation:d3}) with galactocentric distance 
can be well approximated by second-order polynomial expressions. The fits to the these data points  
are shown by the dashed lines, The dotted lines show instead the coefficients 
in the simple relation, Eq.~(\ref{equation:d1}). 

The simple relation between abundance and surface brightness is 
\begin{equation}
12+\log {\rm (O/H)}  =   8.070 (\pm 0.011) + 0.212 (\pm 0.006) \, \log (\Sigma_{L_{W1}})
\label{equation:allr1}
\end{equation}
The coefficients in the simple relation were derived using the abundances and surface brightnesses 
at all the considered galactocentric distances $r$ = 0, 0.1$R_{25}$, 0.2$R_{25}$, ...,  1.0$R_{25}$ 
(990 data points). The general parametric OH -- $SB$ relation is 
\begin{eqnarray}
       \begin{array}{lll}
12+\log {\rm (O/H)}    & = &  7.732 + 0.303 \, r/R_{25} + 0.290 \, (r/R_{25})^{2}  \\  
                       & + & (0.288 + 0.120 \, r/R_{25} - 0.139 \, (r/R_{25})^{2}) \, \log (\Sigma_{L_{W1}})_{r} \\  
                       & + & (0.0418 - 0.0323 \, r/R_{25} - 0.0022 \, (r/R_{25})^{2}) \, h_{W1}  \\
                       & - & (0.0122 + 0.0404 \, r/R_{25} + 0.0088 \, (r/R_{25})^{2}) \, T \\
     \end{array}
\label{equation:allr3}
\end{eqnarray}

Panel $d$ in Figure~\ref{figure:a-r} shows the mean deviations from the 
simple relation, i.e., the residuals of Eq.~(\ref{equation:allr1}) using
crosses and around the parametric relation, i.e., the residuals of 
Eq.~(\ref{equation:allr3}) using circles as a 
function of galactocentric distance. 

All the coefficients in the parametric O/H -- $SB$ relation vary with the galactocentric distance. 
Inspection of panel $c$ in Figure~\ref{figure:a-r} shows that the absolute value of 
coefficient $C_{T}$ increases with increasing galactocentric distance.  
The influence of the morphological type on the OH -- $SB$ relation is negligible 
at the centers of galaxies and increases with galactocentric distance. 
In contrast, the value of coefficient $C_{h}$ decreases with increasing of galactocentric distance.  
The influence of the disk scale length on the OH -- $SB$ relation is largest  
at the centers of galaxies and decreases with galactocentric distance. 
Its influence becomes negligible at the isophotal $R_{25}$ radii of the galaxies. 
The two-dimensional relation, O/H=$f({\Sigma}_{L_{W1}},T)$,  
reproduces the observed data at the optical edges of the disks, 
and the two-dimensional relation, O/H=$f({\Sigma}_{L_{W1}},h_{W1})$,  
reproduces the observed data at the centers of the disks as was 
shown above. 

Examination of panel $d$ in Figure~\ref{figure:a-r} shows that the deviation
from 
the parametric relation is smaller by a factor of $\sim$1.4 than that from 
the simple  
relation at any galactocentric distance. The deviations from both the
parametric and simple relations are smallest at a galactocentric distance of
$r$ $\sim$ 0.4$R_{25}$. It should be noted that the coefficient $C_{SB}$ is 
largest at $r$ $\sim$ 0.4$R_{25}$ (panel $b$ in Figure~\ref{figure:a-r}), i.e., 
the dependence between oxygen abundance and surface brightness is strongest 
at this galactocentric distance. It is interesting also to note that 
Zaritsky, Kennicutt and Huchra have suggested to use  the value of the oxygen
abundance at $r=0.4R_{\rm 25}$  as the characteristic
oxygen abundance in a galaxy \citep{Zaritsky1994ApJ420}. 

It was suggested that it is preferable to compare the properties of 
different galaxies not at a galactocentric distance equal to a fixed 
fraction of the optical radius $R_{25}$ 
but at a galactocentric distance equal to a fixed number of 
the disk scale length, in particular, at the effective radius of a galaxy $R_{eff}$ = 1.68$\times$$h$, 
\citep[e.g.][]{Garnett1987ApJ317,Garnett2002ApJ581,RosalesOrtegaetal2012ApJ756,Sanchez2014aph}. 
Therefore we have obtained the parametric OH -- $SB$ relation at a several values of galactocentric distances 
proportional to the disk scale length $h_{W1}$  in each galaxy, $k$\,$h_{W1}$. 
However, this approach has the following problem. 
The galactocentric distance  $k$\,$h$ with $k$ $\ge$ 3 does not always
lie within the optical radius $R_{25}$ 
for our galaxies (see panel $c$ in Figure~\ref{figure:h-f-t}). Therefore we 
can consider the  OH -- $SB$ relation only up to $k$ = 3. Even in this case we  
have to reject several galaxies with $R_{25}$ $<$ 3$h$. This subsample 
includes 82 galaxies.

Figure~\ref{figure:a-er} shows the obtained values for the 
coefficients in the parametric  
OH -- $SB$  relation at different galactocentric distances $k$ = $r$/$h_{W1}$.  
Panel $d$ in Figure~\ref{figure:a-er} shows the deviations from the 
simple relation (crosses) and from the parametric relation (circles) as a 
function of galactocentric distance $k$ = $r$/$h_{W1}$.  A
comparison between Figure~\ref{figure:a-r} and Figure~\ref{figure:a-er} shows that the 
behavior of the coefficients of the parametric O/H -- $SB$ relation with galactocentric distance 
depends on the choice of the galactocentric distance 
(proportional to the optical radius 
or to the disk scale length). The behavior of the coefficient $C_{h}$ 
exhibits the most appreciable variation. 
The value of the coefficient $C_{h}$ decreases with increasing 
galactocentric distance 
if the characteristics of different galaxies are taken at a fixed fraction of the optical radii. 
The value of the coefficient $C_{h}$ does not exhibit an appreciable change with galactocentric distance 
if the properties of different galaxies are taken at a fixed value of the disk scale length.

Examination of panel $d$ in Figure~\ref{figure:a-er} shows that again the deviation from 
the parametric relation is smaller than that from the simple 
relation at any galactocentric distance $k$. The deviations from both the
parametric and simple relations are minimum at a galactocentric distance 
near the effective radius. 
A comparison between panel $d$ in  Figure~\ref{figure:a-r}
and panel $d$ in Figure~\ref{figure:a-er} shows that the deviations from the parametric 
OH -- $SB$ relation constructed for the abundances and surface brightnesses 
at a galactocentric 
distance of $r$ = 0.4$R_{25}$ is close to that for the abundances 
and surface brightnesses at a galactocentric distance of $k$ = 1.68, $R_{eff}$. 
Thus, the OH -- $SB$ relation varies with galactocentric distance 
and from galaxy to galaxy as in the case when the galactocentric distance 
is chosen as fraction of the optical radius or when the galactocentric 
distance is chosen as a given number of disk scale length. 
For the sake of completeness, the galactocentric distances as fractions 
of the optical radius will be considered below.

\subsection{The Z -- SB relations for sample of galaxies with pure exponential disks} 

\begin{figure}
\epsscale{0.75}
\plotone{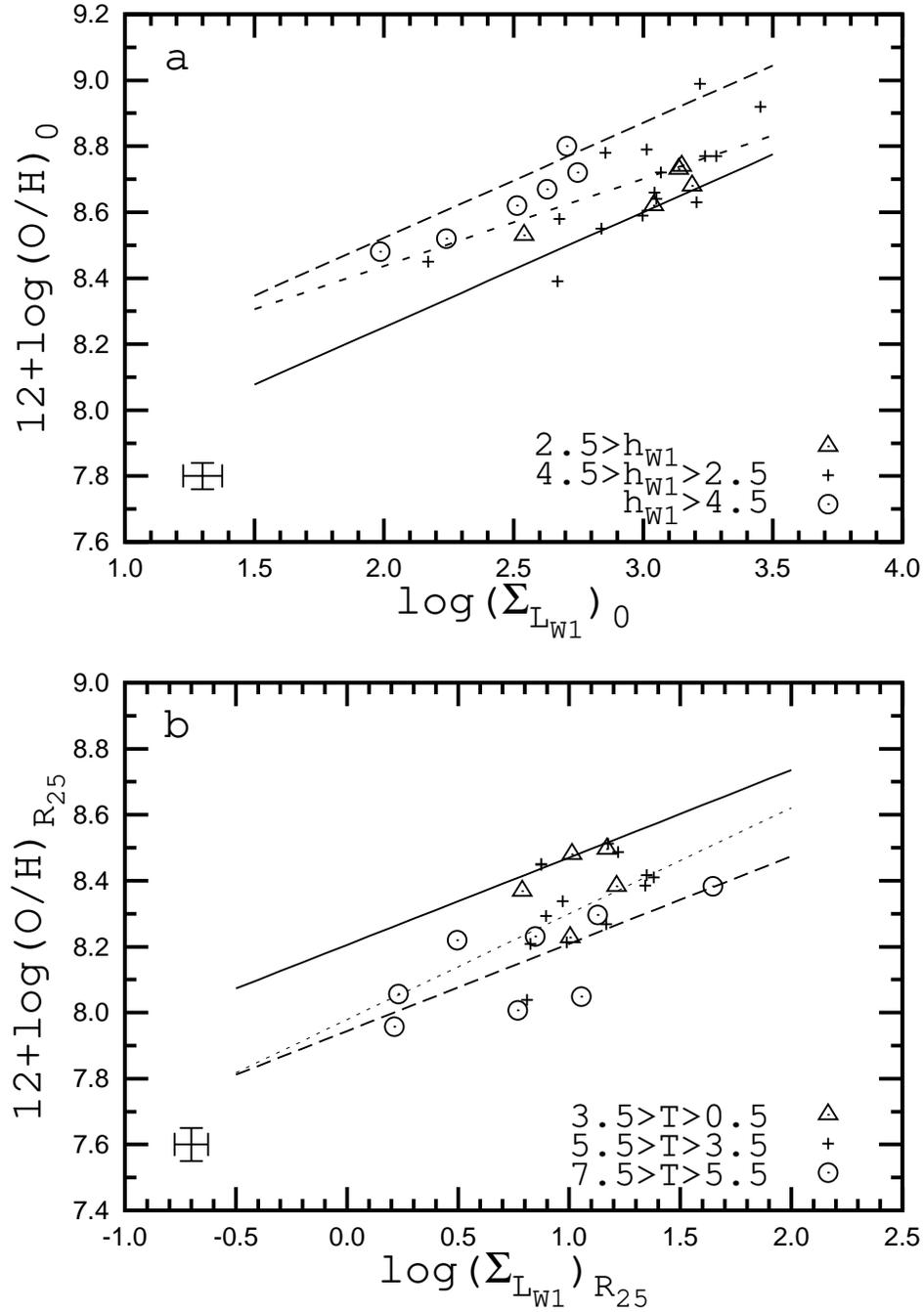}
\caption{
Panel $a$. The same as panel $a$ in  Figure~\ref{figure:oho} but for
galaxies with pure exponential disks.  Panel $b$. The same as panel
$c$ in  Figure~\ref{figure:ohr} but for galaxies with pure exponential
disks. 
\label{figure:ohw05}
}
\end{figure}

For our entire sample of spiral galaxies, we found evidence that the OH -- $SB$ relation 
varies with galactocentric distance and from galaxy to galaxy. 
In general, the four-dimensional   O/H=$f({\Sigma}_{L_{W1}},T,h_{W1},r)$ relation should be used. 
The influence of the morphological type on the OH -- $SB$ relation is negligible 
at the centers of galaxies and increases with galactocentric distance. 
In contrast, the influence of the disk scale length on the OH -- $SB$ relation is largest  
at the centers of galaxies and decreases with galactocentric distance. 
Its influence in fact disappears at the optical edges of galaxies. 
The two-dimensional relation, O/H=$f({\Sigma}_{L_{W1}},T)$,  
reproduces the observed data at the optical edges of the disks, 
and the two-dimensional relation, O/H=$f({\Sigma}_{L_{W1}},h_{W1})$,  
reproduces the observed data at the centers of the disks. 

The parameters of the surface brightness distribution across the disk
obtained through bulge-disk decomposition assuming a pure
exponential for the disk were used for all our galaxies.  However, the
radial surface brightness profiles of only a fraction of galaxies
can be well fitted by a pure exponential while the surface
brightness distribution of the rest of the galaxies is better
described as a broken exponential.  Can the use of pure
exponential disk parameters for galaxies with a broken exponential
disk distort the OH--$SB$ relation and lead to wrong conclusions?
To investigate this point we consider now a subsample of galaxies with
pure exponential disks, with $\sigma_{PED}$ $<$ 0.05 (see
Figure~\ref{figure:s1-s2}).  This subsample contains 26 galaxies. 

We determined the OH--$SB$ relations at the center of the disk and the
mean deviations for this subsample of galaxies.  The coefficients in
the regression equations are listed in Table \ref{table:coefficients}.
The  OH--$SB$ relations for the total sample and for the subsample of
the galaxies with pure exponential disks agree within their
uncertainties (compare the values of the coefficients in  Table
\ref{table:coefficients}).  The mean deviation from the
one-dimensional relation amounts to 0.098, from the two-dimensional relation
O/H=$f({\Sigma}_{L_{W1}}, h_{W1})$ it is 0.081, from the
two-dimensional relation O/H=$f({\Sigma}_{L_{W1}}, f_{B})$ it is
0.096, and from the two-dimensional relation
O/H=$f({\Sigma}_{L_{W1}}, T)$ it is 0.091.  The scatter around all
relations for our subsample of galaxies with pure exponential disks is
slightly lower than that for the complete sample.  But again, the
two-dimensional relation O/H=$f({\Sigma}_{L_{W1}}, h_{W1})$ reproduces
the observed data better than other relations as in the case of the
total sample of galaxies.  Panel $a$ of Figure~\ref{figure:ohw05}
shows the central oxygen abundance as a function of central surface
brightness of the disk in the $W1$ band for galaxies with pure
exponential disks, dividing those galaxies into three groups according
to the value of their disk scale length $h_{W1}$. 

Furthermore, we examine the relations between abundance and surface
brightness at the optical edge of a galaxy's $R_{25}$ isophotal radius
for this subsample of galaxies with pure exponential disks.  The
coefficients in the regression equations are listed in Table
\ref{table:coefficients}.  Again, the  OH--$SB$ relations for the
full sample and for the subsample of galaxies with pure exponential
disks are in agreement within the uncertainties.  We found the mean
deviation from the one-dimensional relation to be 0.124, from the
two-dimensional relation O/H=$f({\Sigma}_{L_{W1}}, h_{W1})$ to amount
to 0.124, from the two-dimensional relation
O/H=$f({\Sigma}_{L_{W1}}, f_{B})$ to be 0.119, and from the
two-dimensional relation O/H=$f({\Sigma}_{L_{W1}}, T)$ to be 0.109.
The values of the mean deviation around the relations for our
subsample of galaxies with pure exponential disks are slightly lower
than the mean deviations for the corresponding relations for the full
sample, and the two-dimensional relation O/H=$f({\Sigma}_{L_{W1}}, T)$
reproduces the observed data better than other relations as in the
case of the complete sample of galaxies.  Panel $b$ of
Figure~\ref{figure:ohw05} shows the oxygen abundance as a function of
surface brightness at the optical edge of galaxies with $\sigma_{1}$
$<$ 0.05, dividing those galaxies into three groups according to their
morphological $T$-type.

\begin{figure}
\epsscale{0.50}
\plotone{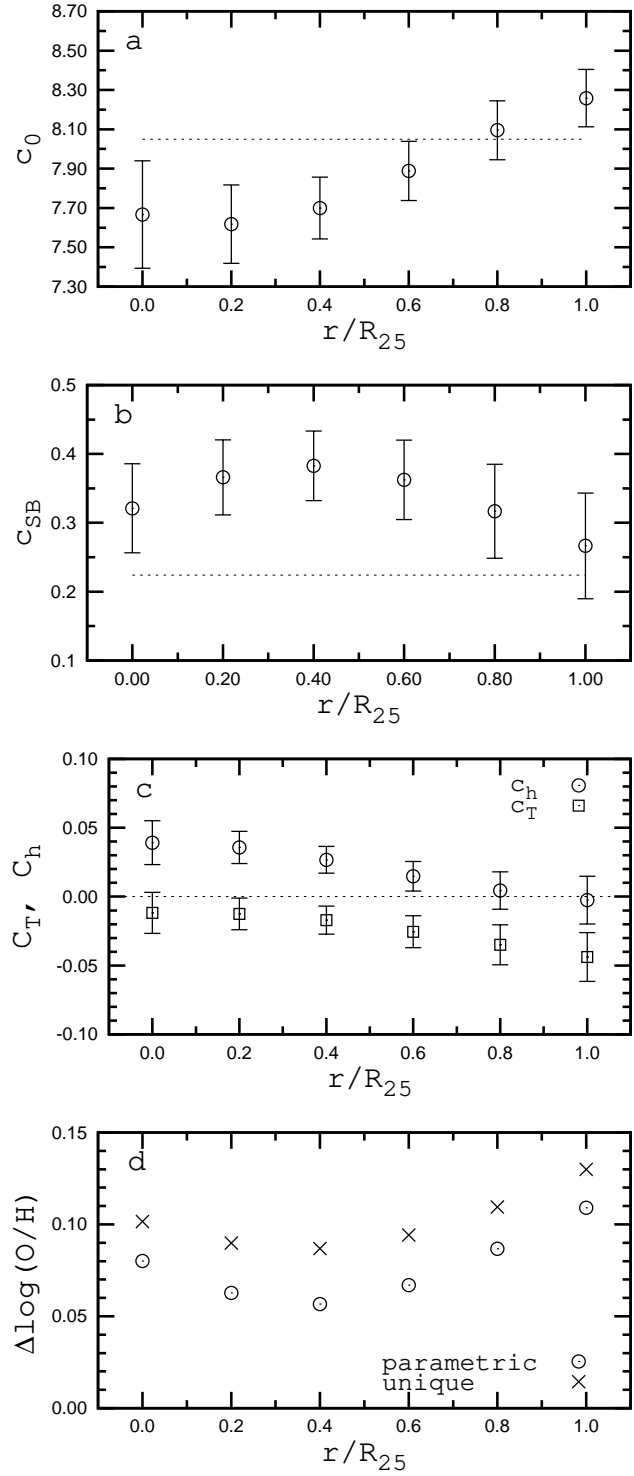}
\caption{
The same as  Figure~\ref{figure:a-r} but for galaxies with pure exponential disks.  
\label{figure:a-rp}
}
\end{figure}

We obtained the parametric relations between abundance and surface brightness 
at different galactocentric distances with a step size of 
$\Delta$$r$ = 0.2$R_{25}$ for the subsample of galaxies with 
pure exponential disks. We also obtained 
the simple OH -- $SB$ relation, which for this subsample of galaxies is 
\begin{equation}
12+\log {\rm (O/H)}  =   8.050 (\pm 0.024) + 0.224 (\pm 0.012) \, \log (\Sigma_{L_{W1}})
\label{equation:purer1}
\end{equation}
Within the uncertainties, this relation agrees with simple relation for 
the entire sample of galaxies,  Eq.~(\ref{equation:allr1}). 

Figure~\ref{figure:a-rp} shows the obtained  coefficients $C_{0}$ (panel $a$),  
$C_{SB}$ (panel $b$), $C_{h}$ (panel $c$, circles) and $C_{T}$ (panel $c$, squares) of the parametric 
relation as a function of galactocentric distance. 
Panel $d$ in Figure~\ref{figure:a-rp} shows the deviations from the 
simple relation (crosses) and from the parametric relation (circles) as a 
function of galactocentric distance. The 
comparison of Figure~\ref{figure:a-r} and Figure~\ref{figure:a-rp} leads to the 
conclusion that the coefficients of the parametric relations 
show a similar general behavior for both the full sample of galaxies and 
the subsample of galaxies with pure exponential disks. 
The residuals of the unique and parametric relations also show a similar general behavior for both 
samples of galaxies. 

Thus, the use of pure exponential disk parameters for galaxies
with a broken exponential disk does not change the general picture.

\subsection{The relations between abundance and surface brightness 
in the $B$ and $K$ bands} 

\begin{figure*}
\epsscale{0.75}
\plotone{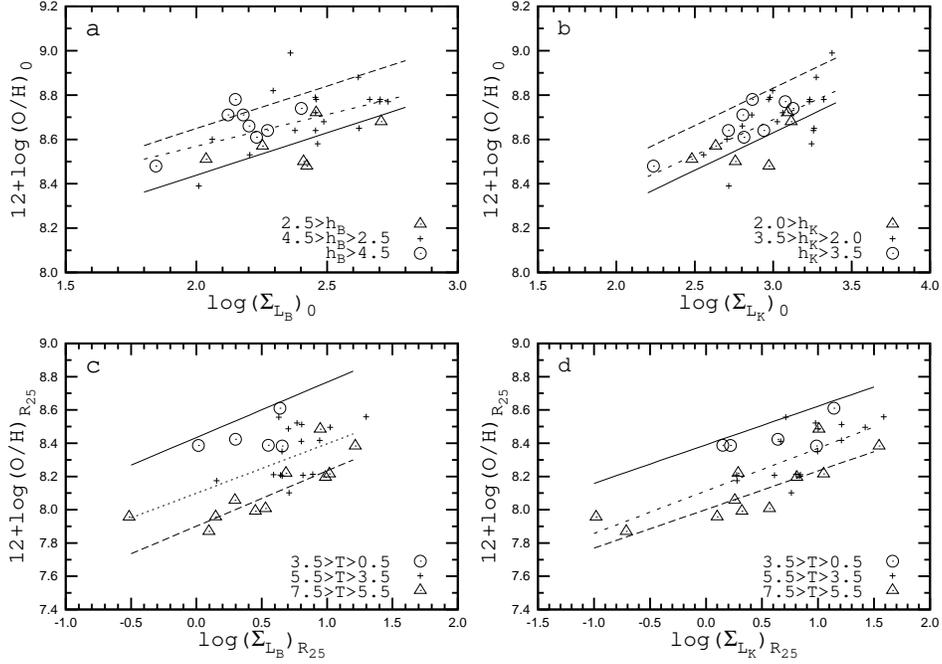}
\caption{
Panels $a$ and $b$ show the central oxygen abundance as a function of
central surface brightness of the disk in the $B$ and $K$ bands.
Panels $c$ and $d$ show this quantity at the optical edge of a galaxy.
The dotted line in each panel shows the one-dimensional relation.
Panel $a$ illustrates the division of galaxies into three subsamples
according to the value of their disk scale length $h_{B}$.  The solid
line corresponds to the parametric relation O/H=$f({\Sigma}_{L_{B}},
h_{B})$ for $h_{B}$ = 1 kpc while the dashed line represents the
parametric relation for $h_{B}$ = 7 kpc.  Panel $b$ shows the same as
panel $a$ but for the $K$ band.  Panel $c$ shows the division of our
sample of galaxies into three subsamples according to the
morphological $T$ type. The solid line is the parametric relation
O/H=$f({\Sigma}_{L_{B}}, T)$ for $T$ = 1, the dashed line is that for
$T$ = 7.  Panel $d$ shows the same as panel $c$ but for the $K$ band. 
\label{figure:ohbk}
}
\end{figure*}

\begin{figure*}
\epsscale{0.75}
\plotone{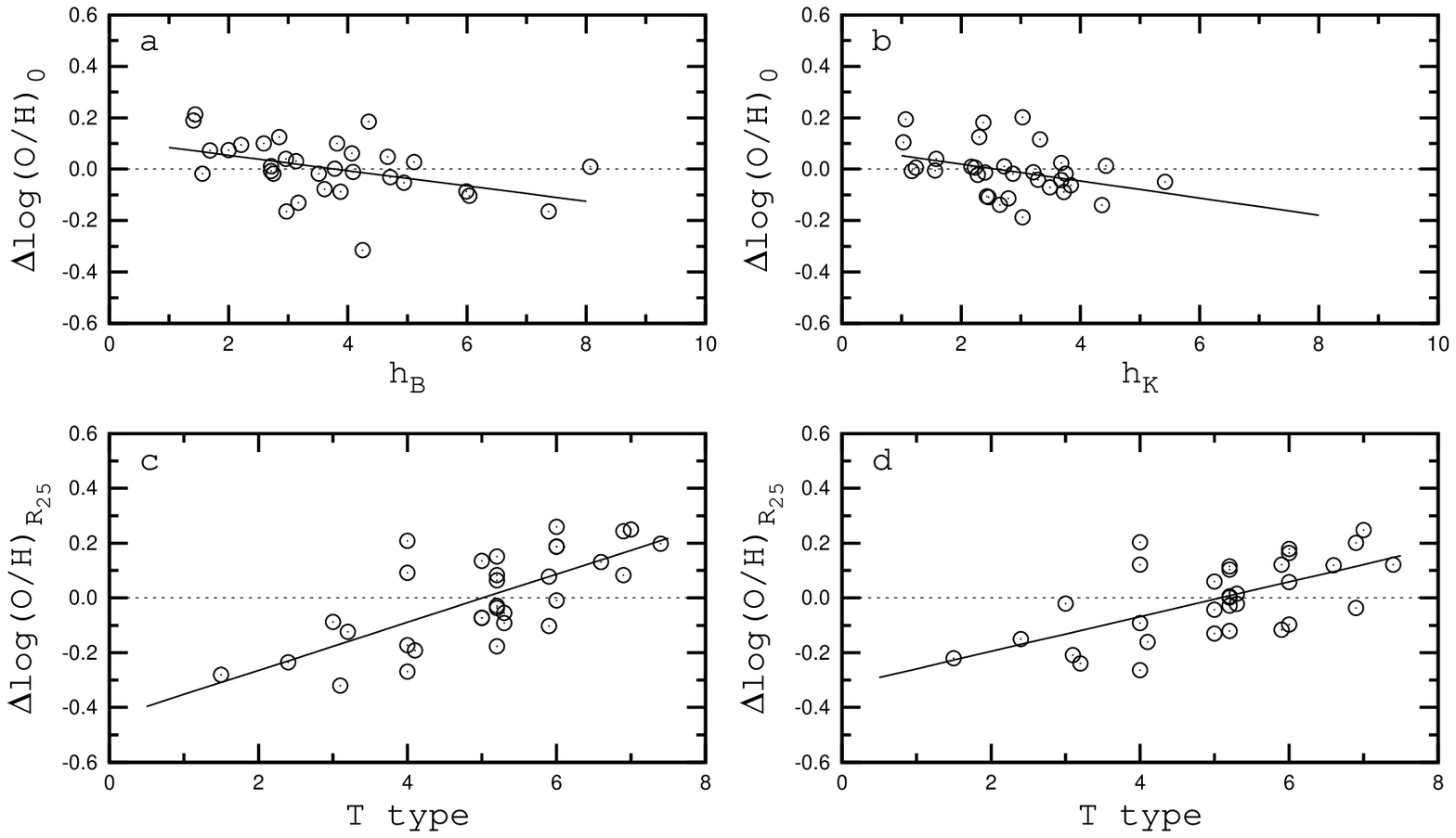}
\caption{
Panel $a$. The residuals of the OH--$SB$ relation for the $B$ band at
the center of the disk as a function of the disk scale length $h_{B}$.
The points indicate the values of the individual galaxies. The solid
lines are linear best fits to those data.  The dotted lines show
zero-lines.  Panel $b$. The same as panel $a$ but for the $K$ band.
Panel $c$. The residuals of the OH--$SB$ relation for the $B$ band at
the optical edge of the disk as a function of the morphological $T$
type.  Panel $d$. The same as panel $c$ but for the $K$ band.
\label{figure:residualbk}
}
\end{figure*}

For 32 galaxies in our sample we have compiled the radial surface
brightness profiles in the photometric $B$ and $K$ bands.  Table
\ref{table:samplebk} lists the central surface brightness of the disk
and the disk scalelength in the $B$ and $K$ bands for each galaxy.
With these data in hand we can examine whether the parametric relation
between abundance and surface brightness reproduces the observed data
better than the one-dimensional relation in the $B$ and $K$ bands.  

The OH--$SB$ relations at the center of the disk in the $B$ and $K$ bands
were obtained in the same way as for the $W1$ band.  The coefficients
in the regression equations are listed in Table
\ref{table:coefficients}.  Panel $a$ of Figure~\ref{figure:ohbk} shows
the central oxygen abundance as a function of central surface
brightness of the disk in the $B$ band.  The galaxies with disk scale
lengths from the three intervals are presented with different symbols.
Panel $a$ of Figure~\ref{figure:residualbk} shows the residuals around
the one-dimensional OH--$SB$ relation as a function of the disk scale
length $h_{B}$.  For the $B$ band, the mean deviation around the
simple relation is 0.110.  The mean deviation around the parametric
relation O/H=$f({\Sigma}_{L_{B}}, h_{B})$ is 0.097.  

Panel $b$ of Figure~\ref{figure:ohbk}  shows the central oxygen
abundance as a function of central surface brightness of the disk in
the $K$ band.  For the $K$ band, we found the mean deviation from 
the one-dimensional relation to be 0.095, and the mean deviation from 
the parametric relation O/H=$f({\Sigma}_{L_{K}}, h_{K})$ to amount to
 0.088.  Panel $b$ of Figure~\ref{figure:residualbk} shows the
residuals of the one-dimensional OH--$SB$ relation as a function
of the disk scale length $h_{K}$. 

We obtained the OH--$SB$ relations at the optical edge of disk in the
same manner.  The coefficients in the regression equations are listed
in Table \ref{table:coefficients}.  Panel $c$ of
Figure~\ref{figure:ohbk} shows the oxygen abundance as a function of
surface brightness of the disk at the isophotal $R_{25}$ radius in the $B$ band.
Panel $c$ of Figure~\ref{figure:residualbk}  shows the residuals
of the one-dimensional OH--$SB$ relation as a function of the
morphological $T$ type.  For the $B$ band, we found the mean deviation
from the simple relation to be 0.167, and the mean deviation
from the parametric relation O/H=$f({\Sigma}_{L_{B}}, h_{B})$ to be
0.115.  

Panel $d$ of Figure~\ref{figure:ohbk} shows the oxygen abundance as a
function of surface brightness of the disk at the  isophotal $R_{25}$ radius
in the $K$ band.  Panel $d$ of Figure~\ref{figure:residualbk}  shows the
residuals from the simple OH--$SB$ relation as a function of the
morphological $T$ type.  For the $K$ band, we found the mean deviation
from the one-dimensional relation to amount to 0.140, and the mean
deviation from the parametric relation O/H=$f({\Sigma}_{L_{K}},
h_{K})$ to be 0.108.  

The comparison between the mean deviation from the one-dimensional
and parametric relations suggests that the parametric OH -- $SB$
relations reproduce the observed data better than the simple relation
also in the $B$ and $K$ bands.  However, the second parameter plays a
more important role in the OH--$SB$ relations at the  isophotal $R_{25}$ radius
of the disk than at the center of the disk for our sample of galaxies
with available surface brightness profiles in the $B$ and $K$ bands.
Figure~\ref{figure:ohbk} and Figure~\ref{figure:residualbk} illustrate
this.  The panels $c$ and $d$ of Figure~\ref{figure:ohbk} for the
values at the optical edge of the disk show clearly that the positions
of galaxies with small values of the $T$ type are, on average, offset
from the positions of galaxies with large values of the $T$ type for
surface brightnesses in both the $B$ and $K$ bands. The shift of
positions of galaxies with short disk scale lengths in the panels $a$
and $b$ (for the values at the center of the disk) relative to the
positions of galaxies with large disk scale lengths is less distinct.
The correlation between the residuals of the one-dimensional OH--$SB$ 
relations at the  isophotal $R_{25}$ radius of the disk and the morhological
$T$ type is much more pronounced (panels $c$ and $d$ of
Figure~\ref{figure:residualbk}) than the correlation between the
residuals of the simple OH--$SB$ relations at the center of the
disk and the disk scale length (panels $a$ and $b$ of
Figure~\ref{figure:residualbk}). 

Thus, the general properties of the abundance -- surface brightness
relations in the $B$ and $K$ bands are similar to those in the $W1$
band. The parametric relation between abundance and surface brightness
reproduces the observed data better than the one-dimensional relation
both at the center of the disk and at optical edge of the disk. For
our sample of galaxies with available surface brightness profiles in
the $B$ and $K$ bands, the second  parameter seems to play a more
important role in the abundance -- surface brightness relation at the
optical edge of the disk than at the center of the disk.

\section{Discussion and conclusions} 

\begin{figure*}
\epsscale{0.75}
\plotone{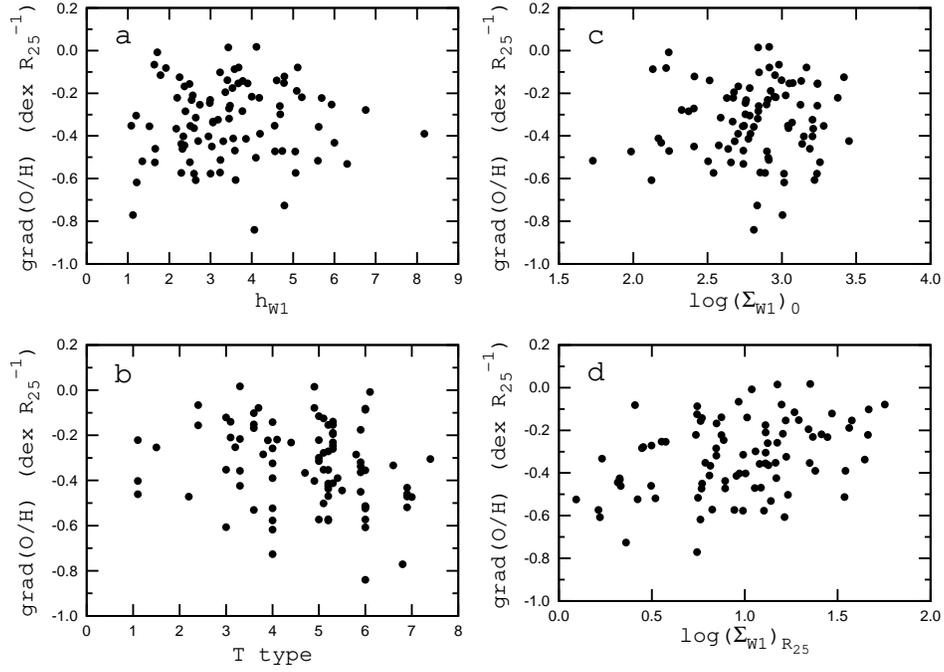}
\caption{
The radial oxygen abundance gradients in units of dex $R_{25}^{-1}$ as
a function of disk scale length $h_{W1}$ (panel $a$), morphological
$T$ type (panel $b$), central disk surface brightness in the $W1$ band
$(\Sigma_{W1})_{0}$ (panel $c$), and disk surface brightness at the
optical edge $(\Sigma_{W1})_{R_{25}}$ (panel $d$).  
\label{figure:grad}
}
\end{figure*}

\begin{figure}
\epsscale{0.60}
\plotone{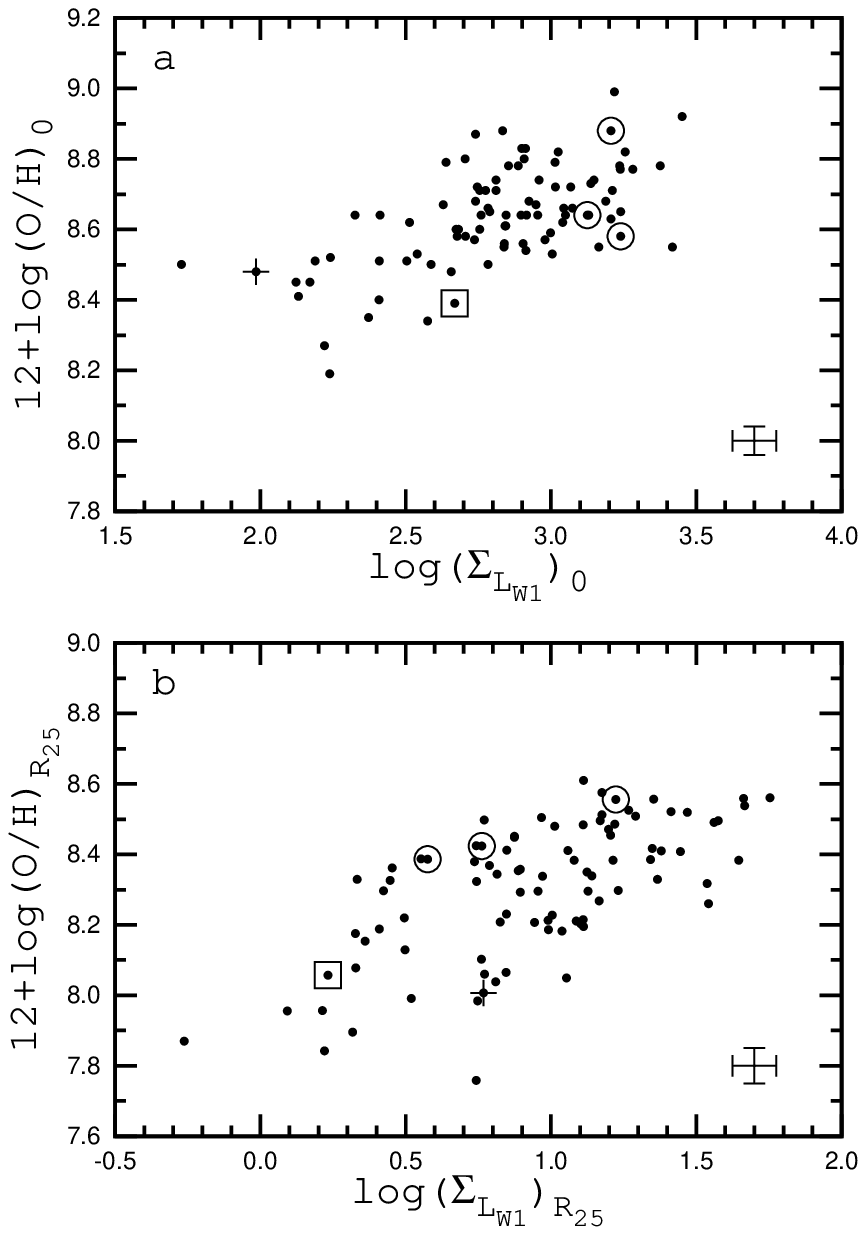}
\caption{
Panel $a$. Central oxygen abundance as a function of central surface
brightness of the disk in the $W1$ band. The open square marks the
positon of the interacting galaxy NGC 4631 with a pure exponential
disk. The open circles denote the interacting galaxies NGC 3031, NGC
3227, and NGC 5194 with broken exponential disks. The plus sign stands
for the possibly interacting galaxy NGC 925 with a pure exponential
disk.  The points indicate other galaxies. 
Typical errors are shown in right lower corner.  
Panel $b$.  The same as panel $a$ but for the optical disk edge.
\label{figure:s-oh-inter}
}
\end{figure}

We found evidence that the parametric relation between abundance and
surface brightness reproduces the observed data better than the simple
relation both at the center of the disk and at the optical edge of the
disk. The second parameter is not unique: the disk scale length should
be used as a second parameter in the relation between abundance and
surface brightness at the center of the disk while the morphological
$T$-type should be used as a second parameter in the relation between
abundance and surface brightness at optical edge of the disk. 

It is difficult to compare directly the numerical values obtained here
and in previous studies because different methods for abundance
determinations were used in different works.  Also, different studies
considered surface brightnesses in different photometric bands. Hence
we will compare the conclusions on the qualitative behavior of
characteristics common to different studies. The most commonly
considered characteristic in these different studies seems to be the
radial abundance gradient. 

\citet{VilaCostas1992MNRAS259} and \citet{Martin1994ApJ424} have
concluded that the global abundance gradients of spiral galaxies with
a barred structure are in general shallower than gradients of
non-barred galaxies. \citet{VilaCostas1992MNRAS259} have also
concluded that non-barred galaxies show a correlation of gradient
slope with morphological type.  \citet{Zaritsky1994ApJ420} found that
the slopes of the radial abundance gradients, when expressed in units
of dex per isophotal radius, do not significantly correlate with either
luminosity or Hubble type.  \citet{Sanchez2014aph} concluded that disk
galaxies show a common or characteristic gradient in the oxygen
abundance (the use of different normalization radii only changes the
numerical values of the common slopes). There is no a significant
dependence on the morphological type, presence or absence of bars,
absolute magnitude and/or stellar mass.  We found in Paper 1
\citep{Paper1} that the radial oxygen abundance gradients (in units of
dex $R_{25}^{-1}$) within the optical radius do not show any
correlation with the morphological type and galaxy radius.  Thus, the
results of these previous studies suggest that the slopes of the
radial abundance gradients, when expressed in units of dex/isophotal
radius, do not significantly correlate with other properties of
galaxies.  Do our present data and results agree with this conclusion? 
 
The radial oxygen gradient in units of dex $R_{25}^{-1}$ is the
difference between the oxygen abundance at the optical edge of a
galaxy's $R_{25}$ isophotal radius and the central oxygen abundance.
Our present results suggest that the former value depends on the
surface brightnesses at the optical edge of a galaxy and on its
morphological type. The latter parameter depends on the surface
brightness at the center of the disk and on the disk scale length.
Thus the radial oxygen gradient in units of dex $R_{25}^{-1}$ should
depend on four parameters.  (To convert the radial oxygen gradient in
units of dex $R_{25}^{-1}$ to the physical gradient in units of dex
kpc$^{-1}$ one needs to know the galaxy radius. The radial oxygen
gradient in units of dex kpc$^{-1}$ should depend on five parameters.)
Since all those parameters vary from galaxy to galaxy, the correlation
between the radial abundance gradient and any individual parameter can
be masked. Figure~\ref{figure:grad} shows the radial oxygen abundance
gradients in units of dex $R_{25}^{-1}$ as a function of disk scale
length $h_{W1}$ (panel $a$), morphological $T$ type (panel $b$),
central disk surface brightness in the $W1$ band $(\Sigma_{W1})_{0}$
(panel $c$), and disk surface brightness at the optical edge
$(\Sigma_{W1})_{R_{25}}$ (panel $d$).  As can be seen in
Figure~\ref{figure:grad}, the radial oxygen abundance gradient does
not significantly correlate with any individual parameter.  This
confirms the statement of \citet{Zaritsky1994ApJ420} that the lack of
a correlation between the gradients and the macroscopic properties of
late-type galaxies may suggest that the relationship between these
parameters is more complex than a simple correlation.

The finding of a flattening of the gradient in barred galaxies in
early studies \citep{VilaCostas1992MNRAS259,Martin1994ApJ424} is not
confirmed by the recent investigation of \citet{Sanchez2014aph}.
Nonetheless, we do not consider this question here for the following
reason. The presence of a bar seems to be a property of a large
fraction of galaxies: around two thirds of the galaxies from our sample of
galaxies are barred galaxies according to the {\sc leda} data base.
The large-scale mixing of the interstellar gas across the disks of
barred spiral galaxies and the radial redistribution of elements seems
to be controlled not so much by the presence or absence of a bar but
primarily by its properties (relative length of the bar, ellipticity,
and bar mass fraction compared to the mass of other components)
\citep{Martin1994ApJ424}.  We plan to investigate the influence of the
bar parameters on the position of a galaxy in the OH -- $SB$ diagram
in a future paper.
For this purpose, we will construct the two-dimensional distribution
of the surface brightness in the $W1$ band for our sample of galaxies.
The results will be reported elsewere. 

It is known that the radial abundance gradients in interacting or
currently merging galaxies are shallower than the gradients of
isolated galaxies
\citep{Rupke2010ApJ723,Rich2012ApJ753,Sanchez2014aph}.  Merger
simulations predict that interacting galaxies should show depressed
nuclear gas metallicities compared to isolated star-forming galaxies
due to the interaction-induced infall of metal-poor gas, and the
radial metallicity of the disk should flatten due to radial mixing of
gas \citep{Rupke2010ApJ710}.   There are five interacting systems
among our galaxies with available surface brightness profiles in both
the $B$ and $W1$ bands. \\  
-- \citet{Pisano1998AJ115} carried out a detailed investigation of NGC
925.  They found that this object is a very asymmetric galaxy (both
morphologically and kinematically).  They detected a $\sim$10$^7$
$M_{\sun}$  H\,{\sc i} cloud apparently interacting with NGC 925.
While the interaction between NGC 925 and the cloud may be responsible
for the observed asymmetries, given the weakness of the interaction
they concluded that NGC 925 has suffered other gravitational
encounters over the past few Gyr (and the H\,{\sc i} cloud might
possibly have been left over from a previous encounter).  It should be
noted that \citet{Rupke2010ApJ723} have considered NGC 925 as a member
of their control sample of galaxies but they did not include this
galaxy in the interacting galaxies sample.  In other words, the
interaction in the  NGC 925 is not beyond question.  \\ 
-- The galaxy NGC 3031 (M81) is one of the best local examples of an
interacting galaxy \citep{Solima2010AA516}.  NGC 3031 is the principal
galaxy of a group that also contains NGC 3034 (M82), NGC 2976, NGC
3037, IC 2574, as well as a large number of dwarf galaxies.  There is
solid evidence of strong interactions among several galaxies of this
group \citep[e.g.,][]{Karachentsev2002A+A383,Makarova2002A+A396}.  
Atomic hydrogen observations revealed remnants of tidal
bridges (with a large H\,{\sc i} clouds) connecting the galaxies
\citep[][and references therein]{Yun1994Nature372,Boyce2001ApJ560}. \\  
-- The galaxy NGC 3227 is a nearby Seyfert galaxy that is interacting
with its dwarf elliptical companion NGC 3226
\citep{Mundell1995MNRAS277,Mundell2004ApJ614}.  Neutral hydrogen
observations revealed tidal tails extending up to $\sim$100 kpc
\citep{Mundell1995MNRAS277}. These authors also discovered a
$\sim$10$^8$ $M_{\sun}$ cloud of H\,{\sc i} close to, but physically
and kinematically distinct from, the galactic disk of NGC 3227.
\citet{Mundell1995MNRAS277} suggested that this cloud might be a third
galaxy in the interacting system accreted by NGC 3227 or that it might
be gas stripped from the disk of NGC 3227, possibly making it a
candidate tidal dwarf galaxy.  \\ 
-- The nearly edge-on Sc galaxy NGC 4631 even has two prominent
companions: the dwarf elliptical NGC 4627 and the edge-on spiral NGC
4656 \citep{Rand1994AA285}. \\ 
-- The galaxy NGC 5194 (=M 51) is known to be interacting with its
companion NGC 5195. A broad  H\,{\sc i} tail extends across a
projected length of 90 kpc \citep{Roots1990AJ100}. 

Two out of those galaxies can be considered as galaxies with pure
exponential disks: NGC 925 with $\sigma_{PED}$ = 0.039 and NGC 4631
with $\sigma_{PED}$ = 0.044.  Panel $a$ of Figure~\ref{figure:grad}
shows the central oxygen abundance as a function of central surface
brightness of the disk in the $W1$ band. The square marks the position
of the interacting galaxy NGC 4631, which has a pure exponential disk.
The circles stand for the interacting galaxies NGC 3031, NGC 3227, and
NGC 5194, which all have broken exponential disks. The plus sign
indicates the possibly interacting galaxy NGC 925, which has a pure
exponential disk.  The points denote other galaxies.  Inspection of
panel $a$ of Figure~\ref{figure:grad} shows that the galaxies NGC
3031, NGC 3227, and NGC 4631 lie near the lower envelope of the
OH--$SB$ diagram, i.e., their central oxygen abundances are lower
than the average value for a given surface brightness. In contrast,
the galaxy NGC 5194 is located  near the upper envelope of the
OH--$SB$ diagram.  It should be noted that we consider not the
locally measured oxygen abundances at the center and at the optical
edge of the disk but the intersect values. Therefore we test the
influence of the interaction on the global gradient but not on the
local oxygen abundances.  

Panel $b$ of Figure~\ref{figure:grad} shows the oxygen abundance as a
function of central surface brightness of the disk in the $W1$ band at
the optical disk edge.  Panel $b$ of Figure~\ref{figure:grad} suggests
that the four interacting galaxies located near the upper envelope of
the OH--$SB$ diagram, i.e., their oxygen abundances at the optical
edge are higher than the mean value for a given surface brightness.
Thus, judging from this small sample, the central oxygen abundances
and the abundances at the optical disk edge in interacting galaxies
seem to follow the prediction of \citet{Rupke2010ApJ710} (with the
exception of the central oxygen abundance in the galaxy NGC 5194). 

Our results show that the bulge does not play an important role in the
present-day oxygen abundance of the disk even at its center. This may
support the suggestion that the bulge is formed at the early epoch of
a galaxy's evolution.  The present-day location of a system in the
$\mu$--O/H diagram is governed by its evolution in the recent past,
and is only weakly dependent on its evolution on long time scales
\citep{Pilyugin1998AA336,Dalcanton2007ApJ658}.  Furthermore, the
observed oxygen abundance in a galaxy is defined not only by its
astration level or gas mass fraction $\mu$, but also by the mass
exchange between a galaxy and its environment \citep{Pagel1997book}.
The infall of low-metallicity gas onto the disk can compensate and
mask the contribution of the bulge stars to the enrichment of the gas
in heavy elements.

In summary, the main results of the present study are the following. 

We constructed radial surface brightness profiles of 95 nearby
late-type galaxies in the infrared $W1$ band using the photometric
maps obtained by the {\it Wide-field Infrared Survey Explorer (WISE)}
project \citep{Wright2010AJ140}.  The characteristics of the bulge and
the disk for each galaxy were obtained through bulge-disk
decomposition assuming both pure and broken exponential disks.  The
characteristics of the bulge and the disk of 32 galaxies were
additionally obtained in the optical $B$ and infrared $K$ bands using
published surface brightness profiles or photometric maps. 

These data coupled with the oxygen abundances presented in our
previous paper were used to examine the relations between the oxygen
abundance and the disk surface brightness at different fractions of the 
optical radius $R_{25}$.  We found evidence that the OH -- $SB$ relation 
depends on the galactocentric distance (taken as a fraction of the optical radius 
$R_{25}$) and on the properties of a galaxy:  the disk scale length and the 
morphological $T$-type. 
The influence of the parameters on the OH -- $SB$ relation 
varies with galactocentric 
distance. The influence of the morphological type on the OH -- $SB$ relation 
is negligible at the centers of the galaxies and increases with galactocentric 
distance. On the other hand, the influence of the disk scale length on the 
OH -- $SB$ relation is largest at the centers of the galaxies and decreases 
with galactocentric distance, disappearing at the optical edges of galaxies. 
The two-dimensional relations can be used to reproduce the observed 
data at the optical edges of the disks and at the centers of the disks. 
However, the second parameter is not unique: the disk scale length should 
be used as a second parameter in the OH -- $SB$ relation at the center of the
disk while the morphological $T$-type should be used as a second
parameter in the relation at optical edge of the disk.

The deviations from the parametric reation are lower by a factor of $\sim$1.4 
than the deviations from the simple, one-dimensional relation. 
The deviations from both parametric and simple relations vary with 
galactocentric distance and are smallest at a galactocentric distance equal 
to 0.4 times the optical radius $R_{25}$.

We have also constructed the OH -- $SB$ relation for abundances and surface brightnesses 
taken at a galactocentric distance equal to a fixed number of the disk scale length 
$h_{W1}$. Again the OH -- $SB$ relation varies with galactocentric distance 
and from galaxy to galaxy as in the case when the galactocentric distance 
is chosen as fraction of the optical $R_{25}$ radius. 
The deviations from the parametric relation are lower than the deviations from the  
one-dimensional relation at any galactocentric distance. 
The deviations from both one-dimensional and parametric 
relations are minimum when the abundances and surface brightnesses are taken 
at a galactocentric distance  near the effective radius of the galaxies.

The relations between oxygen abundance and disk surface brightness in 
the optical $B$ and infrared $K$ bands at the center of the
disk and at optical edge of the disk are also considered.  
The general properties of the abundance -- surface brightness relations are similar 
for all the three considered bands.  




\acknowledgments

We are grateful to the referee for his or her constructive comments. \\
L.S.P., E.K.G., and I.A.Z.\ acknowledge support from 
the Sonderforschungsbereich (SFB 881) on the ``The Milky Way System''
(especially subproject A5), which is funded by the German Research
Foundation (DFG).  L.S.P., and I.A.Z. thank the Astronomisches Rechen-Institut
at Heidelberg University where this investigation was carried out for
the hospitality.  A.Y.K.\ acknowledges the support from the National
Research Foundation (NRF) of South Africa.  I.A.Z.\ acknowledges the
special support by the NASU under the Main Astronomical Observatory
GRID/GPU computing cluster project. \\ 
We thank  Z-Y.~Li and  L.C.~Ho for supporting us with $B$ profiles 
of some galaxies from their catalog in numerical form. \\

This research made use of Montage, funded by the National Aeronautics 
and Space Administration's Earth Science Technology Office, Computational 
Technnologies Project, under Cooperative Agreement Number NCC5-626 between 
NASA and the California Institute of Technology. The code is maintained by 
the NASA/IPAC Infrared Science Archive.

\clearpage

\begin{deluxetable}{lccccccccccc}
\tablewidth{0pt}
\rotate
\tablecaption{\label{table:samplew} 
The parameters of the surface brightness profiles of the late-type galaxies 
of our sample in the $W1$ band obtained through bulge-disk decomposition.
}
\tablehead{
\colhead{Galaxy}                     &
\colhead{Incl.$^{a}$}                & 
\colhead{P.A$^{b}$.}                       &
\colhead{log$(\Sigma_{L})_{e}$$^{c}$}    &
\colhead{$r_{e}$$^{d}$}                     &
\colhead{n$^{e}$}                          &
\colhead{log$(\Sigma_{L})_{0}$$^{f}$}   &
\colhead{$h$$^{g}$}                    &
\colhead{$f_b$$^{h}$}                    &
\colhead{log$L_{W1}$$^{i}$}   &
\colhead{$\sigma_{PED}$$^{j}$}                &
\colhead{$\sigma_{BED}$$^{k}$}                 \\ 
}
\startdata

NGC    12 &  36 & 124 &       &       &      &  2.84 &  3.43 &  0.00 &  10.66 &  0.0513 &  0.0159 \\ 
NGC    99 &  28 & 115 &       &       &      &  2.84 &  3.45 &  0.00 &  10.56 &  0.0420 &  0.0260 \\ 
NGC   224 &  70 &  42 &  3.57 &  1.28 &  2.0 &  2.75 &  4.55 &  0.59 &  11.22 &  0.0328 &  0.0300 \\ 
NGC   234 &  28 &  98 &  2.32 &  1.10 &  1.0 &  3.24 &  3.67 &  0.02 &  11.13 &  0.0505 &  0.0214 \\ 
NGC   253 &  74 &  51 &  4.29 &  0.37 &  1.0 &  3.42 &  2.25 &  0.28 &  11.06 &  0.0780 &  0.0460 \\ 
NGC   300 &  41 & 117 &  0.90 &  0.78 &  2.7 &  2.50 &  1.35 &  0.03 &   9.55 &  0.0728 &  0.0552 \\ 
NGC   450 &  50 &  80 &  1.96 &  0.87 &  1.0 &  2.37 &  2.39 &  0.10 &   9.91 &  0.0819 &  0.0368 \\ 
NGC   493 &  72 &  56 &       &       &      &  2.41 &  3.03 &  0.00 &  10.12 &  0.0503 &  0.0302 \\ 
NGC   575 &  25 &  45 &  1.89 &  0.93 &  1.0 &  2.68 &  3.36 &  0.03 &  10.45 &  0.0326 &  0.0277 \\ 
NGC   628 &  32 &   6 &  2.81 &  0.71 &  1.0 &  2.85 &  3.23 &  0.08 &  10.68 &  0.0409 &  0.0336 \\ 
NGC   783 &  30 &   9 &  2.68 &  2.34 &  1.0 &  2.92 &  5.09 &  0.22 &  11.16 &  0.0563 &  0.0203 \\ 
NGC   925 &  65 & 109 &  2.18 &  1.94 &  1.0 &  1.99 &  5.05 &  0.36 &  10.27 &  0.0387 &  0.0339 \\ 
NGC  1055 &  64 & 103 &  3.24 &  1.92 &  1.0 &  2.51 &  4.60 &  0.68 &  11.06 &  0.0494 &  0.0467 \\ 
NGC  1058 &  23 &  94 &  2.39 &  0.18 &  1.0 &  3.04 &  1.08 &  0.01 &   9.88 &  0.0489 &  0.0275 \\ 
NGC  1068 &  31 &  88 &  4.11 &  0.86 &  1.5 &  2.41 &  4.79 &  0.85 &  11.20 &  0.1547 &  0.1026 \\ 
NGC  1090 &  61 & 102 &  2.10 &  1.27 &  1.0 &  2.84 &  3.77 &  0.04 &  10.80 &  0.0532 &  0.0351 \\ 
NGC  1097 &  44 & 124 &  3.75 &  1.26 &  1.0 &  2.81 &  5.62 &  0.48 &  11.35 &  0.1288 &  0.1003 \\ 
NGC  1232 &  35 &  69 &  2.70 &  0.96 &  1.0 &  2.89 &  5.06 &  0.04 &  11.09 &  0.0530 &  0.0369 \\ 
NGC  1365 &  45 &  13 &  4.05 &  0.99 &  1.1 &  2.90 &  5.93 &  0.45 &  11.49 &  0.1367 &  0.0960 \\ 
NGC  1512 &  48 &  56 &  3.19 &  0.66 &  1.2 &  2.64 &  2.32 &  0.38 &  10.36 &  0.1300 &  0.0640 \\ 
NGC  1598 &  46 & 135 &       &       &      &  3.15 &  2.34 &  0.00 &  10.67 &  0.0481 &  0.0451 \\ 
NGC  1637 &  41 &  34 &  3.14 &  0.39 &  1.0 &  2.95 &  1.79 &  0.13 &  10.27 &  0.1364 &  0.0382 \\ 
NGC  1642 &  30 & 144 &  2.69 &  2.44 &  1.0 &  3.05 &  3.90 &  0.26 &  11.13 &  0.0241 &  0.0195 \\ 
NGC  1672 &  30 & 170 &  3.84 &  0.67 &  1.0 &  2.95 &  4.00 &  0.32 &  11.07 &  0.0737 &  0.0709 \\ 
NGC  2336 &  55 &   5 &  2.93 &  1.69 &  1.0 &  2.71 &  8.18 &  0.15 &  11.30 &  0.0449 &  0.0347 \\ 
NGC  2403 &  57 & 122 &  2.30 &  0.90 &  1.1 &  2.66 &  1.65 &  0.21 &   9.98 &  0.0707 &  0.0616 \\ 
NGC  2441 &  26 &  38 &  2.20 &  0.44 &  1.0 &  2.91 &  4.11 &  0.01 &  10.88 &  0.1356 &  0.0178 \\ 
NGC  2442 &  32 &  99 &  3.51 &  0.69 &  1.0 &  2.92 &  5.11 &  0.15 &  11.08 &  0.0839 &  0.0511 \\ 
NGC  2541 &  59 & 166 &  1.72 &  1.16 &  1.0 &  2.12 &  2.64 &  0.13 &   9.80 &  0.0765 &  0.0534 \\ 
NGC  2805 &  35 &  14 &  2.09 &  1.83 &  1.0 &  2.19 &  6.00 &  0.13 &  10.57 &  0.0618 &  0.0565 \\ 
NGC  2835 &  52 &   5 &  1.73 &  1.18 &  1.7 &  2.59 &  2.64 &  0.07 &  10.20 &  0.0530 &  0.0428 \\ 
NGC  2841 &  63 & 150 &  3.73 &  0.95 &  1.0 &  3.22 &  3.61 &  0.31 &  11.27 &  0.0427 &  0.0248 \\ 
NGC  2903 &  61 &  18 &  3.68 &  0.54 &  1.0 &  3.26 &  2.50 &  0.19 &  10.94 &  0.0857 &  0.0651 \\ 
NGC  2997 &  45 &  98 &  3.39 &  0.58 &  1.0 &  2.91 &  4.10 &  0.11 &  10.94 &  0.1360 &  0.0474 \\ 
NGC  3023 &  47 &  88 &  1.98 &  1.13 &  1.0 &  2.58 &  2.37 &  0.10 &  10.15 &  0.1182 &  0.0466 \\ 
NGC  3031 &  57 & 153 &  3.79 &  0.68 &  1.9 &  3.24 &  2.49 &  0.41 &  11.05 &  0.0769 &  0.0633 \\ 
NGC  3184 &  23 & 148 &  2.69 &  0.53 &  1.0 &  2.78 &  3.53 &  0.04 &  10.65 &  0.1043 &  0.0356 \\ 
NGC  3198 &  68 &  38 &  2.72 &  0.81 &  1.0 &  2.68 &  3.31 &  0.11 &  10.56 &  0.0609 &  0.0400 \\ 
NGC  3227 &  41 & 144 &  3.88 &  0.63 &  1.0 &  3.13 &  2.74 &  0.37 &  10.99 &  0.1478 &  0.0728 \\ 
NGC  3319 &  57 &  40 &  1.99 &  1.42 &  1.0 &  1.73 &  5.60 &  0.25 &   9.97 &  0.0665 &  0.0462 \\ 
NGC  3344 &  26 & 152 &  3.20 &  0.21 &  1.0 &  3.02 &  1.21 &  0.08 &  10.00 &  0.0585 &  0.0518 \\ 
NGC  3351 &  43 &  10 &  3.62 &  0.56 &  1.0 &  3.02 &  2.57 &  0.28 &  10.75 &  0.0762 &  0.0617 \\ 
NGC  3359 &  57 & 174 &  2.41 &  1.27 &  1.0 &  2.41 &  3.45 &  0.22 &  10.36 &  0.0630 &  0.0575 \\ 
NGC  3621 &  61 & 165 &  2.11 &  0.41 &  1.0 &  3.19 &  1.66 &  0.01 &  10.42 &  0.0306 &  0.0285 \\ 
NGC  3631 &  27 & 115 &  3.19 &  0.97 &  1.0 &  2.77 &  3.86 &  0.25 &  10.84 &  0.1742 &  0.0487 \\ 
NGC  3718 &  66 & 178 &  3.46 &  1.06 &  1.0 &  2.67 &  4.18 &  0.44 &  10.94 &  0.0563 &  0.0239 \\ 
NGC  3820 &  36 &  25 &       &       &      &  2.85 &  3.23 &  0.00 &  10.54 &  0.0546 &  0.0163 \\ 
NGC  3893 &  45 & 175 &  2.95 &  1.25 &  1.0 &  3.14 &  2.29 &  0.28 &  10.78 &  0.0432 &  0.0386 \\ 
NGC  3938 &  27 &   3 &  2.74 &  0.81 &  1.0 &  3.01 &  3.00 &  0.07 &  10.78 &  0.0358 &  0.0327 \\ 
NGC  4030 &  36 &  38 &  3.25 &  2.15 &  1.0 &  3.24 &  3.47 &  0.44 &  11.35 &  0.0215 &  0.0184 \\ 
NGC  4088 &  65 &  54 &  2.97 &  0.51 &  1.0 &  3.21 &  2.17 &  0.06 &  10.70 &  0.1469 &  0.0386 \\ 
NGC  4109 &  39 & 153 &       &       &      &  3.21 &  2.95 &  0.00 &  10.93 &  0.0463 &  0.0275 \\ 
NGC  4254 &  41 &  57 &  3.08 &  0.96 &  1.0 &  3.28 &  2.51 &  0.16 &  10.92 &  0.0277 &  0.0247 \\ 
NGC  4303 &  33 & 117 &  3.61 &  0.44 &  1.0 &  3.24 &  2.60 &  0.12 &  10.90 &  0.0716 &  0.0447 \\ 
NGC  4321 &  35 &  24 &  3.43 &  0.83 &  1.0 &  2.96 &  5.21 &  0.14 &  11.20 &  0.1256 &  0.0753 \\ 
NGC  4395 &  47 & 135 &  1.58 &  0.46 &  2.8 &  1.77 &  2.91 &  0.06 &   9.41 &  0.0678 &  0.0567 \\ 
NGC  4501 &  60 & 145 &  3.45 &  0.70 &  1.0 &  3.45 &  2.70 &  0.12 &  11.15 &  0.0500 &  0.0309 \\ 
NGC  4535 &  39 &  16 &  3.13 &  0.64 &  1.0 &  2.75 &  4.69 &  0.09 &  10.89 &  0.1329 &  0.0605 \\ 
NGC  4559 &  67 & 143 &  2.17 &  1.67 &  1.1 &  2.54 &  2.29 &  0.32 &  10.21 &  0.0354 &  0.0281 \\ 
NGC  4625 &  26 & 165 &  1.71 &  0.15 &  1.0 &  3.06 &  0.56 &  0.01 &   9.34 &  0.1305 &  0.0430 \\ 
NGC  4631 &  74 &  82 &  3.21 &  1.55 &  1.0 &  2.67 &  3.06 &  0.64 &  10.87 &  0.0443 &  0.0278 \\ 
NGC  4651 &  47 &  77 &  3.01 &  2.35 &  1.0 &  2.74 &  3.59 &  0.63 &  11.03 &  0.0617 &  0.0446 \\ 
NGC  4654 &  55 & 130 &  2.73 &  0.94 &  1.0 &  3.04 &  2.60 &  0.12 &  10.70 &  0.0431 &  0.0299 \\ 
NGC  4656 &  79 &  34 &  2.05 &  2.84 &  1.0 &  0.95 &  6.35 &  0.86 &  10.10 &  0.0355 &  0.0266 \\ 
NGC  4713 &  35 &  91 &       &       &      &  3.00 &  1.12 &  0.00 &   9.88 &  0.1566 &  0.0328 \\ 
NGC  4725 &  53 &  37 &  3.59 &  0.85 &  1.0 &  2.90 &  4.56 &  0.26 &  11.12 &  0.0710 &  0.0611 \\ 
NGC  4736 &  36 & 100 &  3.96 &  0.50 &  1.9 &  2.98 &  1.64 &  0.70 &  10.71 &  0.1407 &  0.0872 \\ 
NGC  5033 &  63 & 173 &  3.51 &  2.05 &  1.2 &  2.33 &  6.76 &  0.76 &  11.37 &  0.0693 &  0.0504 \\ 
NGC  5068 &  25 &  94 &  2.18 &  0.43 &  1.0 &  2.74 &  1.52 &  0.04 &   9.87 &  0.0665 &  0.0194 \\ 
NGC  5194 &  50 &  28 &  3.46 &  0.82 &  1.0 &  3.21 &  3.18 &  0.19 &  11.08 &  0.1402 &  0.0554 \\ 
NGC  5236 &  18 &  29 &  4.11 &  0.24 &  1.0 &  3.38 &  2.19 &  0.12 &  10.87 &  0.0872 &  0.0679 \\ 
NGC  5248 &  45 & 123 &  3.40 &  1.17 &  1.0 &  3.13 &  3.78 &  0.26 &  11.20 &  0.1089 &  0.0743 \\ 
NGC  5457 &  38 &  66 &  2.91 &  0.45 &  1.0 &  2.81 &  4.06 &  0.03 &  10.84 &  0.0697 &  0.0587 \\ 
NGC  5474 &  28 & 103 &  2.17 &  0.53 &  1.0 &  2.24 &  1.71 &  0.17 &   9.47 &  0.1153 &  0.0546 \\ 
NGC  5668 &  30 & 109 &  2.19 &  2.54 &  1.0 &  2.24 &  4.74 &  0.39 &  10.48 &  0.0416 &  0.0320 \\ 
NGC  6384 &  48 &  33 &  3.00 &  1.87 &  1.3 &  2.74 &  6.31 &  0.28 &  11.22 &  0.0564 &  0.0328 \\ 
NGC  6691 &  28 &  62 &  2.18 &  1.44 &  1.0 &  3.07 &  4.78 &  0.02 &  11.20 &  0.0663 &  0.0198 \\ 
NGC  6744 &  46 &  15 &  3.33 &  0.63 &  1.0 &  2.83 &  4.79 &  0.10 &  11.03 &  0.0589 &  0.0276 \\ 
NGC  6946 &  49 &  61 &  3.76 &  0.30 &  1.2 &  3.07 &  3.06 &  0.10 &  10.81 &  0.0451 &  0.0419 \\ 
NGC  7331 &  65 & 167 &  3.56 &  2.58 &  1.4 &  2.63 &  5.69 &  0.81 &  11.62 &  0.0403 &  0.0338 \\ 
NGC  7495 &  22 &  92 &       &       &      &  2.84 &  4.68 &  0.00 &  10.94 &  0.0737 &  0.0356 \\ 
NGC  7529 &  26 & 146 &       &       &      &  2.76 &  2.54 &  0.00 &  10.27 &  0.0981 &  0.0201 \\ 
NGC  7678 &  33 & 169 &  2.96 &  0.82 &  1.0 &  3.17 &  3.68 &  0.06 &  11.10 &  0.1586 &  0.0399 \\ 
NGC  7793 &  50 &  98 &  2.48 &  0.20 &  1.0 &  2.78 &  1.20 &  0.03 &   9.71 &  0.0581 &  0.0301 \\ 
IC    193 &  37 & 162 &       &       &      &  3.00 &  3.41 &  0.00 &  10.84 &  0.0412 &  0.0095 \\ 
IC    208 &  26 & 171 &       &       &      &  2.71 &  2.37 &  0.00 &  10.22 &  0.1042 &  0.0354 \\ 
IC    342 &  31 &  37 &  3.67 &  0.25 &  2.2 &  2.91 &  3.24 &  0.10 &  10.70 &  0.0993 &  0.0831 \\ 
IC   1132 &  22 &  25 &       &       &      &  2.79 &  4.20 &  0.00 &  10.73 &  0.0688 &  0.0153 \\ 
PGC 45195 &  44 &  63 &  0.84 &  1.54 &  1.0 &  1.77 &  4.91 &  0.03 &   9.82 &  0.0538 &  0.0442 \\ 
UGC  1087 &  23 & 137 &       &       &      &  2.75 &  2.98 &  0.00 &  10.47 &  0.0638 &  0.0146 \\ 
UGC  3701 &  27 & 107 &  1.49 &  1.32 &  1.0 &  2.13 &  3.58 &  0.07 &   9.98 &  0.0769 &  0.0412 \\ 
UGC  4107 &  22 & 134 &       &       &      &  2.90 &  2.99 &  0.00 &  10.59 &  0.0525 &  0.0268 \\ 
UGC  9837 &  31 & 131 &  1.57 &  2.35 &  1.0 &  2.17 &  3.55 &  0.20 &  10.08 &  0.0273 &  0.0215 \\ 
UGC 10445 &  46 & 131 &  1.36 &  0.77 &  1.0 &  2.22 &  1.92 &  0.04 &   9.57 &  0.0597 &  0.0433 \\ 
UGC 12709 &  43 & 148 &  1.04 &  3.83 &  1.0 &  1.49 &  5.96 &  0.27 &   9.85 &  0.0372 &  0.0356 \\ 
\enddata
\tablenotetext{a}{ galaxy inclination ($W1$ band)} 
\tablenotetext{b}{ position angle of the major axis ($W1$ band)} 
\tablenotetext{c}{ {logarithm} of the bulge surface brightness at the effective radius $r_{e}$  
                  in the $W1$ band in $L_{\sun}$ pc$^{-2}$}  
\tablenotetext{d}{ bulge effective radius in kpc} 
\tablenotetext{e}{ shape parameter in the general S\'{e}rsic profile}  
\tablenotetext{f}{ {logarithm} of the central surface brightness of 
                  the disk in the $W1$ band in $L_{\sun}$ pc$^{-2}$}  
\tablenotetext{g}{ disk scale length in the $W1$ band in kpc}  
\tablenotetext{h}{ bulge contribution to the galaxy luminosity in the $W1$ band} 
\tablenotetext{i}{ galaxy luminosity in the $W1$ band} 
\tablenotetext{j}{ mean deviation in the surface brightness profile fitting 
                  through bulge-to-disk decomposition 
                   assuming a pure exponential for the disk}
\tablenotetext{k}{ mean deviation in the surface brightness profile fitting 
                  through bulge-to-disk decomposition 
                   assuming a broken exponential for the disk}
\end{deluxetable}

\clearpage

\begin{deluxetable}{ccccccc}
\tablewidth{0pt}
\tablecaption{\label{table:samplebk} 
Characteristics of the disks in the $B$ and $K$ bands of 
the late-type galaxies of our sample.
}
\tablehead{
\colhead{Galaxy}                     & 
\colhead{log($\Sigma_{L_{B}}$)$_0$$^{a}$}    &
\colhead{$h_B$$^{b}$}                      & 
\colhead{ref$^{c}$}                        &
\colhead{log($\Sigma_{L_{K}}$)$_{0}$$^{d}$}  &
\colhead{$h_K$$^{e}$}                      &  
\colhead{ref$^{f}$}                     
}
\startdata
NGC  234 &    2.62 &  3.82 &   1 &  3.26 &  3.32 &   1 \\ 
NGC  300 &    2.04 &  1.69 &   2 &  2.48 &  1.24 &   3 \\ 
NGC  598 &    2.10 &  1.99 &   4 &  2.64 &  1.34 &   5 \\ 
NGC  628 &    2.46 &  3.61 &   6 &  2.97 &  2.43 &   3 \\ 
NGC  925 &    1.85 &  4.67 &   7 &  2.24 &  3.68 &   8 \\ 
NGC 1232 &    2.15 &  7.37 &   9 &  2.87 &  4.36 &   3 \\ 
NGC 1365 &    2.27 &  8.07 &  10 &  2.71 &  5.42 &   3 \\ 
NGC 1642 &    2.46 &  4.07 &   1 &  2.94 &  3.68 &   1 \\ 
NGC 2403 &    2.42 &  1.44 &  11 &  2.97 &  1.07 &   8 \\ 
NGC 2841 &    2.36 &  4.25 &   8 &  3.37 &  3.03 &   8 \\ 
NGC 3031 &    2.46 &  2.85 &  12 &  3.25 &  2.37 &   8 \\ 
NGC 3184 &    2.20 &  4.71 &   8 &  2.80 &  3.29 &   8 \\ 
NGC 3198 &    2.06 &  4.09 &   8 &  2.71 &  3.21 &   8 \\ 
NGC 3227 &    2.38 &  2.96 &  12 &  3.26 &  2.30 &  13 \\ 
NGC 3344 &    2.46 &  1.56 &  13 &  3.09 &  1.17 &  13 \\ 
NGC 3351 &    2.29 &  2.97 &   8 &  3.00 &  2.65 &   8 \\ 
NGC 3621 &    2.71 &  2.21 &   9 &  3.11 &  1.58 &   8 \\ 
NGC 3938 &    2.45 &  3.88 &   8 &  2.98 &  2.79 &   8 \\ 
NGC 4030 &    2.70 &  3.78 &  13 &  3.08 &  3.84 &  13 \\ 
NGC 4254 &    2.73 &  2.71 &   8 &  3.23 &  2.40 &   8 \\ 
NGC 4303 &    2.70 &  2.71 &  12 &  3.32 &  2.23 &   8 \\ 
NGC 4321 &    2.40 &  4.94 &   8 &  3.13 &  3.75 &   8 \\ 
NGC 4535 &    2.12 &  6.04 &  12 &  2.81 &  3.72 &   3 \\ 
NGC 4559 &    2.20 &  2.59 &  12 &  2.56 &  2.17 &   8 \\ 
NGC 4631 &    2.01 &  4.35 &   8 &  2.72 &  3.03 &   8 \\ 
NGC 4651 &    2.49 &  3.13 &   1 &  3.03 &  2.72 &   1 \\ 
NGC 5055 &    2.14 &  4.20 &   8 &  3.29 &  2.38 &   8 \\ 
NGC 5068 &    2.25 &  2.00 &   9 &  2.63 &  1.56 &   3 \\ 
NGC 5194 &    2.62 &  3.17 &   8 &  3.27 &  2.46 &   8 \\ 
NGC 5236 &    2.66 &  2.75 &  14 &  3.23 &  2.28 &   3 \\ 
NGC 5248 &    2.42 &  4.41 &  12 &  3.14 &  3.97 &   3 \\ 
NGC 5457 &    2.18 &  5.99 &  15 &  2.86 &  3.49 &   3 \\ 
NGC 6946 &    2.46 &  3.51 &  13 &  3.06 &  2.87 &  13 \\ 
NGC 7495 &    2.23 &  5.11 &   1 &  2.81 &  4.43 &   1 \\ 
NGC 7793 &    2.41 &  1.41 &   2 &  2.76 &  1.03 &   3 \\ 
\enddata
\tablenotetext{a}{{Logarithm} of the central surface brightness of 
                  the disk in the $B$ band in $L_{\sun}$ pc$^{-2}$}  
\tablenotetext{a}{Disk scale length in the $B$ band in kpc}  
\tablenotetext{c}{Reference to the source for the photometric data used for 
                  the determination of the central surface brightness of 
                  the disk and its scale length in the $B$ band}
\tablenotetext{d}{{Logarithm} of the central surface brightness of 
                  the disk in the $K$ band in $L_{\sun}$ pc$^{-2}$}  
\tablenotetext{e}{Disk scale length in the $K$ band in kpc}  
\tablenotetext{f}{Reference to the source for the photometric data used for 
                  the determination of the central surface brightness of 
                  the disk and its scale length in the $K$ band}
\tablerefs{
 (1) \citet{deJong1994AAS106}; 
 (2) \citet{Carignan1985ApJS58}; 
 (3) \citet{Jarrett2003AJ125};
 (4) \citet{deVaucouleurs1959ApJ130}; 
 (5) \citet{Regan1994ApJ434}: 
 (6) \citet{Natali1992AA256};
 (7) \citet{Macri2000ApJS128};
 (8) \citet{Munoz2009ApJ703}; 
 (9) \citet{Li2011ApJS197}; 
(10) \citet{Jorsater1995AJ110}; 
(11) \citet{Okamura1977PASJ29}; 
(12) SDSS; 
(13) \citet{Knapen2004AA426}; 
(14) \citet{Talbot1979ApJ229}; 
(15) \citet{White2003PASP115}
}
\end{deluxetable}

\clearpage

\begin{deluxetable}{cccccccc}
\tablewidth{0pt}
\tablecaption{\label{table:coefficients} 
Values of the coefficients of the regression euqation between abundance and surface brightness   
12+log(O/H)$_{r}$ = C$_{0}$ + C$_{1}$ $\times$ log($\Sigma_{L})_{r}$ + C$_{2}$ $\times$ $\alpha$ 
}
\tablehead{
\colhead{band$^{a}$}                     & 
\colhead{r$^{b}$}                        &
\colhead{$\alpha$$^{c}$}                & 
\colhead{$C_{0}$$^{d}$}                   &
\colhead{$C_{1}$$^{d}$}                   &
\colhead{$C_{2}$$^{d}$}                   &
\colhead{$\sigma$$^{e}$}                 &
\colhead{N$^{f}$}                     
}
\startdata
 $W1$ & 0       &         & 7.91 $\pm$ 0.10 & 0.259 $\pm$ 0.035 &                      & 0.113 & 90 \\ 
 $W1$ & 0       & $h_{W1}$ & 7.61 $\pm$ 0.10 & 0.308 $\pm$ 0.031 &  0.0451 $\pm$ 0.0077 & 0.095 & 90 \\ 
 $W1$ & 0       & $T$     & 8.13 $\pm$ 0.12 & 0.222 $\pm$ 0.036 & -0.0258 $\pm$ 0.0088 & 0.107 & 90 \\ 
 $W1$ & $R_{25}$ &         & 8.01 $\pm$ 0.04 & 0.307 $\pm$ 0.038 &                      & 0.144 & 90 \\ 
 $W1$ & $R_{25}$ & $h_{W1}$ & 7.96 $\pm$ 0.05 & 0.298 $\pm$ 0.038 &  0.0184 $\pm$ 0.0112 & 0.142 & 90 \\ 
 $W1$ & $R_{25}$ & $T$     & 8.34 $\pm$ 0.06 & 0.277 $\pm$ 0.031 & -0.0622 $\pm$ 0.0090 & 0.116 & 90 \\ 
 $W1$ & 0       &         & 7.91 $\pm$ 0.16 & 0.263 $\pm$ 0.055 &                      & 0.098 & 26 \\ 
 $W1$ & 0       & $h_{W1}$ & 7.51 $\pm$ 0.18 & 0.349 $\pm$ 0.054 &  0.0451 $\pm$ 0.0140 & 0.081 & 26 \\ 
 $W1$ & 0       & $T$     & 8.17 $\pm$ 0.20 & 0.223 $\pm$ 0.056 & -0.0287 $\pm$ 0.0145 & 0.091 & 26 \\ 
 $W1$ & $R_{25}$ &         & 7.98 $\pm$ 0.08 & 0.321 $\pm$ 0.078 &                      & 0.124 & 26 \\ 
 $W1$ & $R_{25}$ & $h_{W1}$ & 7.97 $\pm$ 0.10 & 0.318 $\pm$ 0.081 &  0.0039 $\pm$ 0.0189 & 0.124 & 26 \\ 
 $W1$ & $R_{25}$ & $T$     & 8.25 $\pm$ 0.13 & 0.265 $\pm$ 0.074 & -0.0435 $\pm$ 0.0171 & 0.109 & 26 \\ 
 $B$  & 0       &         & 8.00 $\pm$ 0.21 & 0.286 $\pm$ 0.090 &                      & 0.110 & 32 \\ 
 $B$  & 0       & $h_{B}$  & 7.64 $\pm$ 0.23 & 0.383 $\pm$ 0.088 &  0.0351 $\pm$ 0.0124 & 0.097 & 32 \\ 
 $B$  & 0       & $T$     & 8.35 $\pm$ 0.23 & 0.219 $\pm$ 0.085 & -0.0381 $\pm$ 0.0139 & 0.099 & 32 \\ 
 $B$  & $R_{25}$ &         & 8.10 $\pm$ 0.06 & 0.297 $\pm$ 0.083 &                      & 0.167 & 32 \\ 
 $B$  & $R_{25}$ & $h_{B}$  & 8.05 $\pm$ 0.09 & 0.293 $\pm$ 0.084 &  0.0139 $\pm$ 0.0195 & 0.166 & 32 \\ 
 $B$  & $R_{25}$ & $T$     & 8.52 $\pm$ 0.09 & 0.333 $\pm$ 0.059 & -0.0888 $\pm$ 0.0157 & 0.116 & 32 \\ 
 $K$  & 0       &         & 7.72 $\pm$ 0.19 & 0.321 $\pm$ 0.065 &                      & 0.095 & 32 \\ 
 $K$  & 0       & $h_{K}$  & 7.58 $\pm$ 0.20 & 0.338 $\pm$ 0.062 &  0.0336 $\pm$ 0.0161 & 0.088 & 32 \\ 
 $K$  & 0       & $T$     & 7.97 $\pm$ 0.30 & 0.268 $\pm$ 0.082 & -0.0171 $\pm$ 0.0158 & 0.093 & 32 \\ 
 $K$  & $R_{25}$ &         & 8.11 $\pm$ 0.04 & 0.256 $\pm$ 0.046 &                      & 0.140 & 32 \\ 
 $K$  & $R_{25}$ & $h_{K}$  & 8.00 $\pm$ 0.07 & 0.245 $\pm$ 0.044 &  0.0445 $\pm$ 0.0241 & 0.132 & 32 \\ 
 $K$  & $R_{25}$ & $T$     & 8.45 $\pm$ 0.08 & 0.232 $\pm$ 0.037 & -0.0648 $\pm$ 0.0149 & 0.109 & 32 \\ 
\enddata
\tablenotetext{a}{photometric band of the surface brightness} 
\tablenotetext{b}{value of the radius to which the relation corresponds}  
\tablenotetext{c}{identification of the second parameter in the relation} 
\tablenotetext{d}{value of the coefficient in the regression} 
\tablenotetext{e}{mean deviation of the regression} 
\tablenotetext{f}{number of the data points used}
\end{deluxetable}

\end{document}